\documentstyle[aps,floats,psfig,tighten]{revtex}

\begin{document}


\def\ij{\langle i,j\rangle}
\def\suma{\sum_{\bf a}}
\def\lap{{\Delta_a}}
\def\omk{\omega(\vec k)}
\def\omky{\omega(k_y)}
\def\Ak{A_{\vec k}}
\def\epsaa{\epsilon_{AA}}
\def\epsab{\epsilon_{AB}}
\def\epsbb{\epsilon_{BB}}
\def\pasol{p_{\rm sol}^A}
\def\pbsol{p_{\rm sol}^B}
\def\pavap{p_{\rm vap}^A}
\def\pbvap{p_{\rm vap}^B}

\def\begeq{\begin{equation}}
\def\endeq{\end{equation}}
\def\begarr{\begin{eqnarray}}
\def\endarr{\end{eqnarray}}
\def\begvec{\left(\begin{array}{c}}
\def\endvec{\end{array}\right)}
\def\begmat{\left(\begin{array}{cc}}
\def\endmat{\end{array}\right)}

\title{Spinodal decomposition of an ABv model alloy: 
  Patterns at unstable surfaces}

\author{Mathis Plapp}

\address{Physics Department and Center for Interdisciplinary Research
   on Complex Systems,\\ Northeastern University, Boston, MA 02115, USA}

\author{Jean-Fran\c cois Gouyet}

\address{Laboratoire de Physique de la Mati\`ere Condens\'ee, 
   Ecole Polytechnique, 91128 Palaiseau, France}

\date{June 18, 1999}

\maketitle

\begin{abstract}
We develop mean-field kinetic equations for a lattice gas
model of a binary alloy with vacancies (ABv model) in which
diffusion takes place by a vacancy mechanism. These equations are applied
to the study of phase separation of finite portions of an 
unstable mixture immersed in a stable vapor. Due to a larger 
mobility of surface atoms, the most unstable modes of spinodal decomposition 
are localized at the vapor-mixture interface. Simulations show 
checkerboard-like structures at the surface or surface-directed 
spinodal waves. We determine the growth rates of bulk and surface modes
by a linear stability analysis and deduce the relation between the
parameters of the model and the structure and length scale of
the surface patterns. The thickness of the surface patterns is
related to the concentration fluctuations in the initial state.
\end{abstract}

\section{Introduction}
\nobreak
\noindent
The phase separation of alloys and other unstable mixtures (binary
fluids, glasses, polymer blends) has been extensively studied for 
decades \cite{Cahn58,Cahn68,Gunton83,Binder91,Bray94}.
In a typical experiment, a stable mixture is quenched into an unstable
state by a sudden change in an external control parameter, mostly temperature
or pressure. Then, domains of the new equilibrium phases develop and
coarsen, leading to a heterogeneous material. As the mechanical and
transport properties of an alloy may depend 
considerably on its microstructure,
the understanding of the dynamics of domain formation and growth
is of great practical importance. In addition, 
phase separation is an interesting
example of a process where spontaneous pattern formation occurs
during the approach to equilibrium.

In recent years, attention has been drawn to surface effects
on spinodal decomposition, mainly due to interesting experimental 
results on polymer blends 
\cite{Jones91,Wiltzius91,Cumming92,Bruder92,Shi93,Harrison95}. 
Thin films of mixture are placed between glass plates,
or between a substrate and vacuum, and quenched. If the
interactions with the surface favor one of the components, 
this component rapidly segregates to the surface
and triggers phase separation in the bulk: one observes
surface-directed spinodal decomposition, during which 
concentration waves propagate from the
surface into the sample. In some of the experiments, modulations 
parallel to the surface are observed, leading to a
pattern along the surface with a distinct length 
scale \cite{Wiltzius91,Cumming92,Bruder92}.
The surface effects are in competition with
bulk spinodal decomposition, and only a thin layer
is affected by the surface, whereas in the interior of the
sample the usual bulk structures are found. Interesting questions
are then what determines the type of structures found at the
surface, and to predict typical length scales and growth rates
of the patterns, as well as the thickness of the surface layer.

The dynamics of spinodal decomposition near planar substrates has
been investigated using continuum models \cite{Puri94,Frisch95,Fischer97}, 
Monte Carlo simulations \cite{Sagui93}, and mean-field kinetic 
equations \cite{Geng96}. On the other hand, very little is known 
about free surfaces, which can deform during the decomposition
process \cite{Keblinski96}. We have recently found that surface
modes and spinodal waves at free surfaces can be observed in a
simple lattice model \cite{Plapp97c}, and the present paper is devoted
to a detailed study of this phenomenon.

The classical theory for the early stage of spinodal decomposition,
based on out-of-equilibrium thermodynamics, was
proposed by Cahn and Hilliard \cite{Cahn58}. The Cahn-Hilliard
equation is obtained by postulating that the local interdiffusion currents
are proportional to the gradients of the 
local chemical potential, and supposing
that this chemical potential can be derived from a free energy
functional of Ginzburg-Landau type. The proportionality constant is the
atomic mobility, which is a phenomenological parameter in this theory.
For a homogeneous initial state, the Cahn-Hilliard
equation may be linearized around the average composition,
and analyzed in terms of Fourier modes. Long wavelength
perturbations grow exponentially with time, whereas short wavelength
fluctuations are damped by the gradient terms, related to the
surface tension. A typical length scale of the resulting domain pattern
is then given by the wavelength of the fastest growing mode.

It has been recognized that simplified lattice models are valuable 
tools to study the influence of microscopic dynamics on the process of 
phase separation in metallic alloys. They provide a convenient
conceptual framework with a minimum number of parameters.
Such lattice gas models 
assume the existence of a fixed crystal lattice,
the sites of which can be occupied by the atoms
of different species, or by vacancies. A configuration
evolves by atomic jumps from site to site. Many studies have been 
carried out on the kinetic Ising model with Kawasaki spin exchange 
dynamics \cite{Kawasaki72,Rao76,Lebowitz82,Huse86,Amar88}, 
which corresponds to the direct exchange of atoms. 
In most alloys, however, the dominant mechanism of diffusion is a 
vacancy mechanism \cite{Murch91}, and several models of
vacancy-mediated spinodal decomposition have been investigated 
\cite{Yaldram,Fratzl,Frontera,Soisson96}. In our model, the rates
for atomic jumps follow an Arrhenius law.

A standard procedure to investigate the dynamics of such models are
Monte Carlo simulations \cite{Binder91}. 
The advantage of this approach is that the
simulations constitute a genuine realization of the stochastic model,
and all correlations in space and time are preserved. But for the
same reason, an analytic understanding is difficult. Therefore, there
have been numerous attempts to formulate microscopic
kinetic equations which are analytically tractable. These equations are
derived from the microscopic master equation by an approximation of
mean-field type. While the linearized equation of 
Khatchaturyan \cite{Khatchaturyan68} is valid
only near equilibrium, recently several authors have proposed fully
nonlinear kinetic equations for Ising and lattice gas models 
\cite{Penrose90,Martin90,Gouyet93,Chen94a,Vaks95}. These
equations can be cast in the form of generalized Cahn-Hilliard equations,
where the atomic mobility depends explicitly on the details of the
microscopic dynamics. Such equations have been applied successfully
to the study of phase-separation dynamics in binary and ternary
systems \cite{Chen94b,Gouyet95,Dobretsov}, 
and to dendritic growth \cite{Plapp97a}. In particular, kinetic equations
for models with vacancy dynamics have been proposed by 
Chen and Geng \cite{Chen}
and by Puri and Sharma \cite{Puri97}. We will follow the approach 
developed by one of the present authors \cite{Gouyet93}, which allows
to relate the microscopic equations in a particularly straightforward
manner to the equations of out-of equilibrium thermodynamics.

All the studies of vacancy-mediated phase separation we are aware of
start from a homogeneous initial state. We study here the evolution
of droplets of a dense mixture with few vacancies, immersed in a
bath of vacancy-rich ``vapor'' phase. The interesting point in this
situation is that we can have interfaces between an unstable 
mixture and a stable vapor which are ``neutral'', that is without
segregation of one of the components. Such interfaces are
impossible in a binary model like the Ising model, where any
stable phase below the critical temperature favors one of the components.

In our model, the atoms situated
near the surface of the droplets have a larger mobility than bulk
atoms. Therefore, phase separation is fastest at the surface. We
observe different structures at the surface, depending on the
model parameters and the initial composition of the mixture. In
the specially symmetric case of equal interaction energies and
concentrations for both components, we obtain a regularly modulated
surface mode which generates an ordered pattern in a surface layer.
For asymmetrical interaction energies or compositions, 
surface-directed spinodal waves are observed. This difference
can be explained by a competition between surface spinodal
decomposition and surface segregation.
We determine the characteristic length scales and growth rates
of bulk and surface modes by a linear stability analysis of the 
mean-field kinetic equations. 
Our approach allows thus to relate morphological parameters
to the interaction parameters of the microscopic model.

The remainder of this article is organized as follows: in Sec. 2, 
we present the model and the derivation of the mean-field 
kinetic equations. Sec. 3 describes the simulations, which are 
analyzed in Sec. 4: a linear stability analysis is performed 
in the bulk and at the surface. Sec. 5 presents the
discussion of the results.


\section{Model and mean field kinetic equations}
\nobreak
\noindent
We consider a simple cubic lattice in $d$ dimensions, with lattice 
constant $a$ and a total number of $N$ sites. The
sites can be occupied by atoms of two species, 
$A$ and $B$, or by vacancies $v$.
The occupation numbers at site $i$, $n^\alpha_i$, where $\alpha$
denotes $A$, $B$, or $v$, are equal to $1$ if site $i$ is occupied by
species $\alpha$, and zero otherwise. Double occupancy is forbidden, 
which gives the constraint $n_i^A + n_i^B + n_i^v = 1 \,\forall i$.
Hence there are two independent variables 
per site; we choose in the following $n_i^A$
and $n_i^B$. We consider only interactions between nearest neighbor
atoms: the
energy of a configuration ${\cal C} = \{n_i^A,n_i^B, i = 1\ldots N\}$ is
given by the Hamiltonian
\begeq
H({\cal C}) = -\sum_{\ij} 
    \left[\epsaa n_i^A n_j^A +\epsbb n_i^B n_j^B
          \epsab(n_i^A n_j^B + n_i^B n_j^A)\right],
\label{hamilabv}
\endeq
where the sum is over all nearest neighbor pairs $\ij$, and
$\epsilon_{\alpha\beta}$ are the interaction energies between two
atoms occupying two nearest neighbor sites. Vacancies do not interact
with atoms or other vacancies. Unlike in a binary model,
where the phase diagram is completely determined by the
exchange energy $\epsaa+\epsbb-2\epsab$, here the interaction energies are
independent. One of them sets the temperature scale, the other two are free
parameters. It can be shown that Eq. (\ref{hamilabv}) is equivalent to the 
Hamiltonian of a general ternary system (see appendix).

The configuration evolves by jumps of atoms to a neighboring vacant 
site. This is a common diffusion mechanism in metals, and the 
associated energy barrier is usually much lower than for a
direct interchange of atoms. We therefore will completely neglect the latter
process. As appropriate for activated processes, the hopping rates 
are assumed to follow an Arrhenius law. The underlying physical 
picture is that the atoms are trapped in potential wells located 
around the sites of the lattice. Atoms spend most 
of their time near a lattice site, but from time to time they jump over 
the energy barrier between neighboring sites, the
necessary energy being provided by the lattice phonons. We assume
two contributions to the barriers: a constant activation energy 
$U_\alpha$ and the local binding energy, $\partial H / \partial n_i^\alpha$,
which depends on the local configuration. The jump rate from 
site $i$ to $j$ is
\begeq
w_{i\to j}^\alpha = \nu_0^\alpha \exp\left[-{1\over kT}\left(U_\alpha + 
           {\partial H\over \partial n_i^\alpha}\right)\right].
\endeq
Here, $T$ is the temperature (constant throughout the system) 
and $k$ is Boltzmann's constant. The prefactor $\nu_0^\alpha$ is 
related to a vibration frequency of the atom in the well
(at least in the absolute Eyring regime). 
We can absorb the constant $U_\alpha$ in a prefactor which
sets the time scale, $w_0^\alpha = \nu_0^\alpha \exp(-U_\alpha/kT)$. In
principle, $w_0^A$ and $w_0^B$ are different; however, simulation
results indicate that qualitative results are unaffected by
the ratio $w_0^A/w_0^B$ as long as it is not 
too far from unity \cite{Yaldram}.
Therefore, we will take in the following $w_0^A = w_0^B = w_0$. 
The jump rates become
\begeq
w_{i\to j}^\alpha = 
   w_0 \exp\left[-{\epsilon_{\alpha A}\over kT}\sum_a n_{i+a}^A
                 -{\epsilon_{\alpha B}\over kT}\sum_a n_{i+a}^B\right].
\label{jumprates}
\endeq
Here and in the following, summation over $a$ means a sum over all
nearest-neighbor sites.

The jump rates in Eq. (\ref{jumprates}) define a stochastic process. 
To describe its time evolution, one may use the master equation 
for the probability distribution
${\cal P}({\cal C},t)$ of finding configuration $\cal C$ at time $t$:
\begeq
{\partial{\cal P}({\cal C},t)\over\partial t }=
     \sum_{\cal C'}\left({\cal P}({\cal C'},t) W({\cal C',C}) - 
        {\cal P}({\cal C},t) W({\cal C,C'})\right).
\endeq
The transition rates $W({\cal C,C'})$ from configuration $\cal C$ to $\cal C'$ 
are equal to $w_{i\to j}^\alpha$ given by Eq. (\ref{jumprates}) 
if the two configurations differ
only by an exchange of a particle and a vacancy, and zero otherwise.
Given a solution to the master equation, one can formally define 
a time-dependent average for any operator $O(\{n_i^\alpha\})$, function 
of the occupation numbers, by
\begeq
\left\langle O\right\rangle (t) = 
    \sum_{\cal C} O(\{n_i^\alpha\}) {\cal P}({\cal C},t).
\endeq
In particular, we define the time-dependent occupation probabilities of the
sites:
\begeq
p_i^\alpha(t) = \left<n_i^\alpha\right>(t).
\endeq
These probabilities can also be interpreted as local concentrations.
To obtain a kinetic equation for $p_i^\alpha(t)$, one differentiates 
both sides of the above equation with respect to time 
and uses the master equation. The result is a conservation
law for the local occupation probability,
\begeq
{\partial p_i^\alpha\over\partial t} = -\sum_a j_{i\,i+a}^\alpha.
\label{conserv}
\endeq
The currents $j_{ij}^\alpha$ through the link $i,j$ are given by
\begeq
j_{ij}^\alpha = \left<n_i^\alpha(1-n_j^A-n_j^B) w_{i\to j}^\alpha - 
          n_j^\alpha(1-n_i^A-n_i^B) w_{j \to i}^\alpha\right>.
\label{strom}
\endeq
The prefactors of the jump rates assure that the start site is 
occupied by an $\alpha$-atom and the target site is empty.

Up to now, the kinetic equation is equivalent to the complete master
equation. To obtain a closed system of equations for the occupation 
probabilities
$p_i^\alpha$, we make a mean-field approximation and replace 
the occupation numbers in the expressions for the currents by their 
averages $p_i^\alpha$. Clearly, this approximation is drastic:
the resulting equations are a set of coupled deterministic
differential equations. Fluctuations are suppressed,
and the mean-field approximation does not take into account
correlations. Nevertheless, as in the case of 
static mean-field approximations,
one can use this method to get qualitative results on the dynamics
of microscopic models. Similar approximations have been discussed 
by several authors \cite{Penrose90,Martin90,Gouyet93,Chen94a,Vaks95}. 
A possibility to improve systematically the simple mean-field equations
is the use of the path probability method (PPM) devised by
Kikuchi \cite{Kikuchi66}. The equations of the PPM, however, 
are considerably more complicated, and simulations on bulk 
spinodal decomposition have shown that the use of the PPM in the
pair approximation leads to the same qualitative conclusions as
the simple mean-field approximation \cite{Geng96}. Therefore,
we will study in the following only the mean-field case.

We generalize to our ternary model the method developed for 
binary systems in Ref. \cite{Gouyet93}. The equations
for the currents can be written as a product of a prefactor 
$S_{ij}$, symmetric with respect to the interchange of $i$ and $j$,
and the difference of two local terms $C_j$ and $C_i$, 
equivalent to chemical activities:
\begeq
j_{ij}^\alpha = - S_{ij}^\alpha \left(C_j^\alpha - C_i^\alpha\right),
\endeq
with
\begeq
S_{ij}^\alpha = w_0 (1-p_i^A-p_i^B)(1-p_j^A-p_j^B)
\endeq
and
\begeq
C_i^\alpha = {p_i^\alpha\over 1-p_i^A-p_i^B}
   \exp\left[-{\epsilon_{\alpha A}\over kT}\sum_a p_{i+a}^A
                   -{\epsilon_{\alpha B}\over kT}\sum_a p_{i+a}^B\right].
\endeq
This factorization is not unique (see Ref. \cite{Gouyet93} for more details).
Our particular choice makes it straightforward to establish a connection
to the phenomenological equations of out-of-equilibrium thermodynamics. We
define a local chemical potential by
\begeq
\mu_i^\alpha = kT \ln C_i^\alpha.
\endeq
Then, the current becomes
\begeq
j_{ij}^\alpha = - M_{ij}^\alpha \left(\mu_j^\alpha - \mu_i^\alpha\right),
\label{curmob}
\endeq
where the mobility in the link $ij$ is given by
\begeq
M_{ij}^\alpha = S_{ij}^\alpha 
          {C_j^\alpha - C_i^\alpha \over \mu_j^\alpha - \mu_i^\alpha}.
\label{mobil}
\endeq
The explicit expression for the chemical potentials is
\begeq
\mu_i^\alpha = -\epsilon_{\alpha A} \lap p_i^A 
       -\epsilon_{\alpha B}\lap p_i^B
       -z\epsilon_{\alpha A} p_i^A -z\epsilon_{\alpha B} p_i^B
         +kT \ln{p_i^\alpha\over 1-p_i^A-p_i^B},
\label{chempots}
\endeq
where $\lap g_i = \sum_a (g_{i+a} - g_i)$ for
any site-dependent quantity $g$ is the discrete Laplacian,
and $z$ is the coordination number of the lattice.
This expression can also be obtained as the derivative of a
free energy function $F$ with respect to the local occupation,
\begeq
\mu_i^\alpha = \partial F / \partial p_i^\alpha.
\endeq
The free energy function is a discrete analog of a Ginzburg-Landau 
functional:
\begeq
F = \sum_i f(p_i^A,p_i^B,T) + \sum_a\left( {\epsaa\over 4} 
   {(p_{i+a}^A-p_i^A)}^2 + 
   {\epsab\over 2} {(p_{i+a}^A - p_i^A)(p_{i+a}^B - p_i^B)}
   {\epsbb\over 4} {(p_{i+a}^B-p_i^B)}^2\right) 
\label{freen}
\endeq
with a local free energy density
\begin{eqnarray}
f(p^A,p^B,T) & = & -{z\epsaa\over 2}{p^A}^2 - 
           z\epsab p^A p^B - {z\epsbb\over 2}{p^B}^2 \nonumber \\
    & & \mbox{} + kT\left[p^A\ln p^A + p^B\ln p^B + 
           (1-p^A-p^B)\ln(1-p^A-p^B)\right].
\end{eqnarray}
This free energy could have been obtained by a simple 
static mean field approximation of the Hamiltonian. Furthermore, for a 
closed system (no currents crossing the boundaries), $F$ can only 
decrease. This can be seen explicitly by taking its time 
derivative, using (\ref{curmob}) and noticing that the mobility
is always positive:
\begeq
{d F\over dt} = \sum_{i,\alpha}{\partial F\over \partial p_i^\alpha}
                               {d p_i^\alpha\over dt} = 
                -\sum_{\ij,\alpha} M_{ij}^\alpha 
                         \left(\mu_j^\alpha-\mu_i^\alpha\right)^2.
\endeq
Therefore, the dynamics leads to a state which minimizes the static
mean-field free energy. This final
state may be the ground state (global minimum) or a metastable state
(local minimum); we cannot describe nucleation events, unless we
explicitly introduce fluctuations, for example by adding Langevin
noise to the deterministic equations.

Stated in terms of chemical potentials and mobilities, our kinetic
equations have the form of generalized Cahn-Hilliard equations. In contrast
to the phenomenological equations, the mobilities depend on the local
configuration and are related to the details of the microscopic jump
processes, albeit in an approximate manner. This allows in principle
an application of these equations to situations far from equilibrium,
and to situations where the mobility depends strongly of the local
concentration. For the vacancy dynamics considered here, the concentrations
of $A$ and $B$ are independent variables, but their evolution is coupled
by the vacancy field. The mobilities are higher in regions with high
vacancy concentration. To see this, consider the mobilities for a
homogeneous system. For $p_i^A \to \bar p^A$ and $p_i^B \to \bar p^B$, 
Eq. (\ref{mobil}) becomes
\begeq
M^\alpha_{\rm hom}(\bar p^A, \bar p^B) = 
      w_0 \bar p^\alpha (1-\bar p^A-\bar p^B)\,
      \exp\left[-{z\epsilon_{\alpha A}\over kT}\bar p^A - 
               {z\epsilon_{\alpha B}\over kT}\bar p^B\right].
\label{mobhom}
\endeq
This expression has a simple interpretation: the prefactor is the mean-field
probability of finding an $\alpha$-atom and a vacancy on neighboring sites,
and $w_0$ times the exponential term is the mean jump rate.

It is worth noticing that in the mean-field approximation there are no
``off-diagonal terms'' in the mobility matrix: in general, one should
obtain an expression for the currents of the form 
$j^\alpha = -M^{\alpha A}\nabla \mu^A - M^{\alpha B}\nabla \mu^B$. The
reason for these terms missing is probably the suppression of all
correlations in Eq. (\ref{strom})
by the mean-field approximation. In the present context,
this deficiency seems to be of little importance. It should be noted
that even if the off-diagonal mobilities are zero, the same is not true
for the diffusion coefficients, because the chemical potentials
involve the concentrations of both species.

\begin{figure}
\centerline{
 \psfig{file=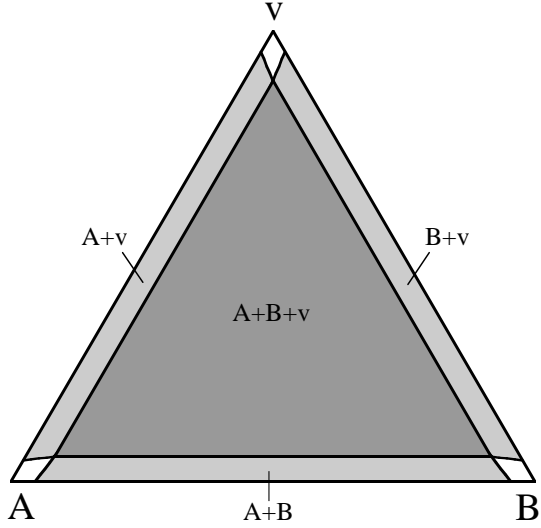,width=0.4\textwidth}}
\caption{Cut through the phase diagram at constant temperature $kT=0.6\epsaa$.
The interaction energies are $\epsaa=\epsbb$ and $\epsab=\epsaa/2$.
White: one phase, light gray: two phase coexistence, dark gray: three
phase coexistence.}
\label{figphasediag}
\end{figure}
From the free energy density, we can obtain the phase diagram. We will
consider only completely attractive models here, where no order-disorder
transitions occur ($\epsaa > 0$, $\epsbb > 0$,
$\epsaa+\epsbb-2\epsab > 0$). The
conditions for phase coexistence are that the chemical potentials and
the grand potential, $\Omega = f - \mu^A p^A - \mu^B p^B$ be equal in the
two (or three) phases. This is equivalent to a ``common
tangent plane'' construction: the free energy density, function of two 
variables, defines a surface in the space $(p^A,p^B,f)$. The (homogeneous)
chemical potentials as functions of $p^A$ and $p^B$ define the orientation
of planes tangent to this surface. The condition of equal grand potential
implies that the points representing two (or more) phases in
equilibrium must lie in the same plane.
Thus we can obtain the phase diagram by constructing
all the planes that are tangent to the free energy surface
in at least two points. Various structures
of phase diagrams can be obtained \cite{Plappth}. For attractive
interactions, quite generally the free energy surface has three
minima for sufficiently low temperatures. Then, there exists exactly
one plane which is tangent in three points: we have three-phase
coexistence. Besides, there are families of double tangent planes
which give the coexistence lines for two-phase coexistence.
For $\epsaa = \epsbb$, our model is equivalent to the 
Blume-Emery-Griffiths model \cite{Blume71}. We will focus 
in this paper on the specific example $\epsab = \epsaa/2$. 
As we then have $\epsaa = \epsbb = \epsaa+\epsbb-2\epsab$, the 
phase diagram is completely symmetric with respect to the 
interchange of any two components and may be calculated
using the analogy with the three state 
Potts model (see appendix).
The phase diagram for $kT/\epsaa = 0.6$ and $z = 4$ (two dimensions) is 
shown in Fig. \ref{figphasediag}. There is a large region where an A-rich, 
a B-rich, and a vacancy-rich ``vapor'' phase coexist. 
In the regions of two-phase coexistence, the concentration of the third 
component is very low.

From formula (\ref{mobhom}) we can immediately deduce that the mobility
in the dilute ``vapor''-phase, where $p^A$ and $p^B$ are small, is much
larger than in the two dense phases. As we have $\epsaa=\epsbb>\epsab$, 
the diffusion of the minority component is always faster than
that of the majority component in the dense phases.


\section{Simulations}
\nobreak
\noindent
The part of the phase diagram which is most appropriate for the 
description of an alloy is the region of AB-coexistence with low 
vacancy concentration. All studies of lattice gas dynamics with 
vacancies we are aware of are limited to this area. We will
investigate the behavior of finite ``droplets'' of such a material 
immersed in a stable ``vapor'', a very natural situation which can
arise for instance when droplets of a liquid mixture in coexistence
with its vapor are rapidly quenched into an unstable state. Evidently,
our model is not adapted to describe diffusion processes in a vapor,
where diffusion does not take place via activated jumps to nearest
neighbor sites. But the important feature of this ``vapor''
phase is that diffusion is much faster than inside the ``solid''.
Moreover, the atomic mobility decreases continuously across
the vapor-mixture interface, and hence at the surface the diffusion
is faster than inside the bulk. As we shall see, 
this induces fast surface modes.

\begin{figure}
\centerline{
  \psfig{file=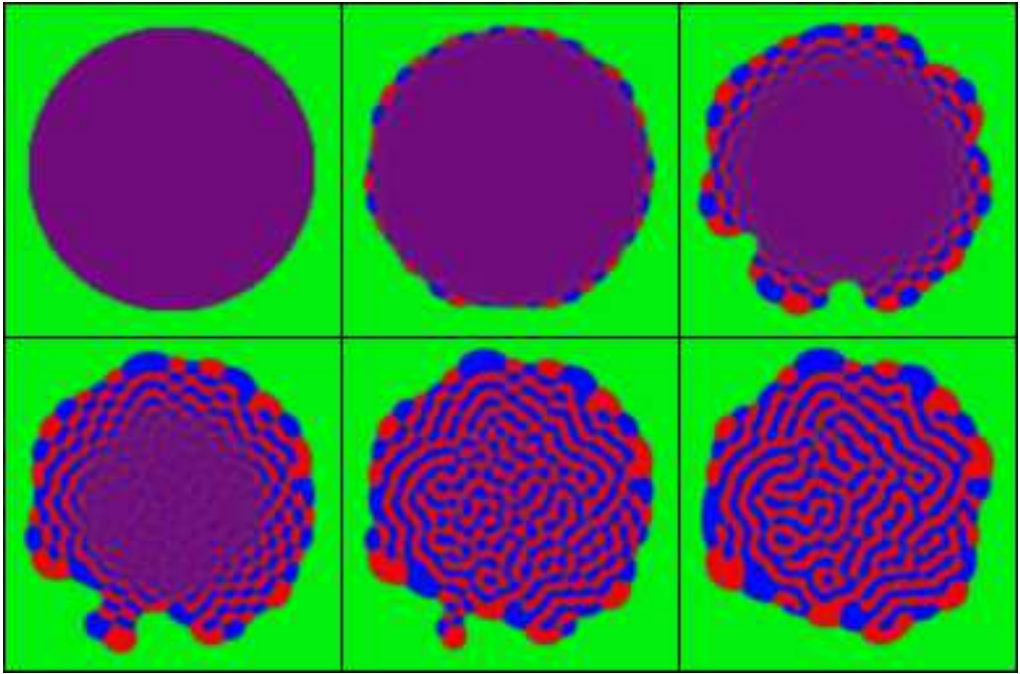,width=0.75\textwidth}}
\medskip
\caption{Snapshot pictures of the decomposition of a droplet on a
$128\times 128$ lattice with periodic boundary conditions. Parameters are: 
$\epsaa = \epsbb = 1$, $\epsab = 0.5$, $kT = 0.5$, $\pasol=\pbsol=0.46464$,
and $\pavap=\pbvap=0.03536$. The intensities of red, blue, and green
are proportional to $p^A$, $p^B$, and $p^v$, respectively.
The pictures are taken at $t = 14$, $24400$,
$49400$, $74400$, $99400$, and $149200$ $w_0^{-1}$.}
\label{figdroplet}
\end{figure}
We integrated the mean-field kinetic equations by an explicit
Euler scheme. The time step 
is limited by the numerical stability of the algorithm. For our
inhomogeneous system, the ``most dangerous'' regions of the simulation 
domain are those where the diffusion is fastest, thus the 
vapor phase. For the diffusion equation with a diffusivity $D_{\rm vap}$,
the maximum time step is $D_{\rm vap}/2da^2$ in $d$ dimensions. 
For the temperatures and vapor
compositions we used, $D_{\rm vap}/a^2w_0$ is slightly
less than unity. We integrated with a maximum time step of $1/4w_0$ in
2D and $1/6w_0$ in 3D without encountering numerical instabilities.
When we started with step functions as initial conditions, the
time step had to be chosen much smaller at the beginning and was then
slowly increased to the maximum value.

As initial state, we chose ``droplets'' (i.~e. circular domains) or slabs of
a mixture with few vacancies (typically some percents) and concentrations
$\pasol$ and $\pbsol$, immersed in a vapor of concentrations 
$\pavap$ and $\pbvap$. To trigger the phase separation, we
added small fluctuations to the initial state. To assure 
mass conservation,
pairs of neighboring sites and a component (A or B) 
were randomly chosen, and the concentrations at the two sites 
were shifted by $+A_0 r$ and $-A_0 r$, respectively,
where $r$ is a random number uniformly distributed between 
$-1$ and $1$. The noise amplitude $A_0$ ranged 
between $10^{-5}$ and $10^{-2}$.
For 2D simulations, we used a lattice 
of size $128\times 128$ with periodic boundary conditions. 
All our simulations were carried out on 
workstations and took from 2 to 20 hours of CPU time. 
We also simulated two 3D-samples on a $32\times32\times64$
lattice; these took up to 100 hours CPU time. 

Let us first discuss critical quenches ($\pasol = \pbsol$).
Snapshot pictures from the time evolution of a droplet are shown 
in Fig. \ref{figdroplet}. Phase separation starts at the surface: 
a fairly regular modulation appears along the mixture-vapor interface. 
This surface mode triggers phase separation in adjacent regions and
propagates into the interior of the sample with a constant velocity, 
leaving behind a checkerboard-like ordered structure. The domains often
coalesce to form stripes. 

\begin{figure}
\centerline{
 \hskip .5in
 \psfig{file=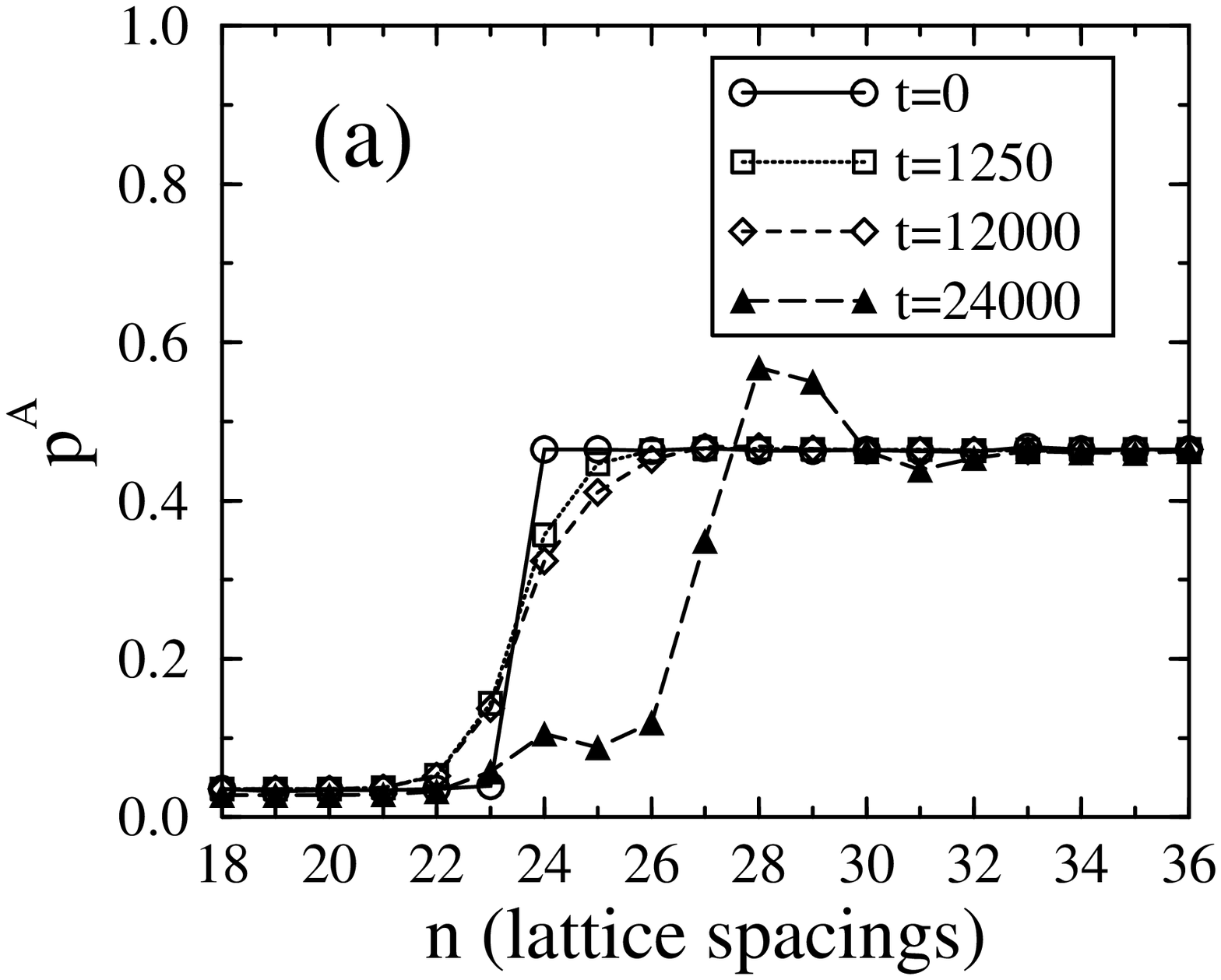,width=0.4\textwidth}\hfill
 \psfig{file=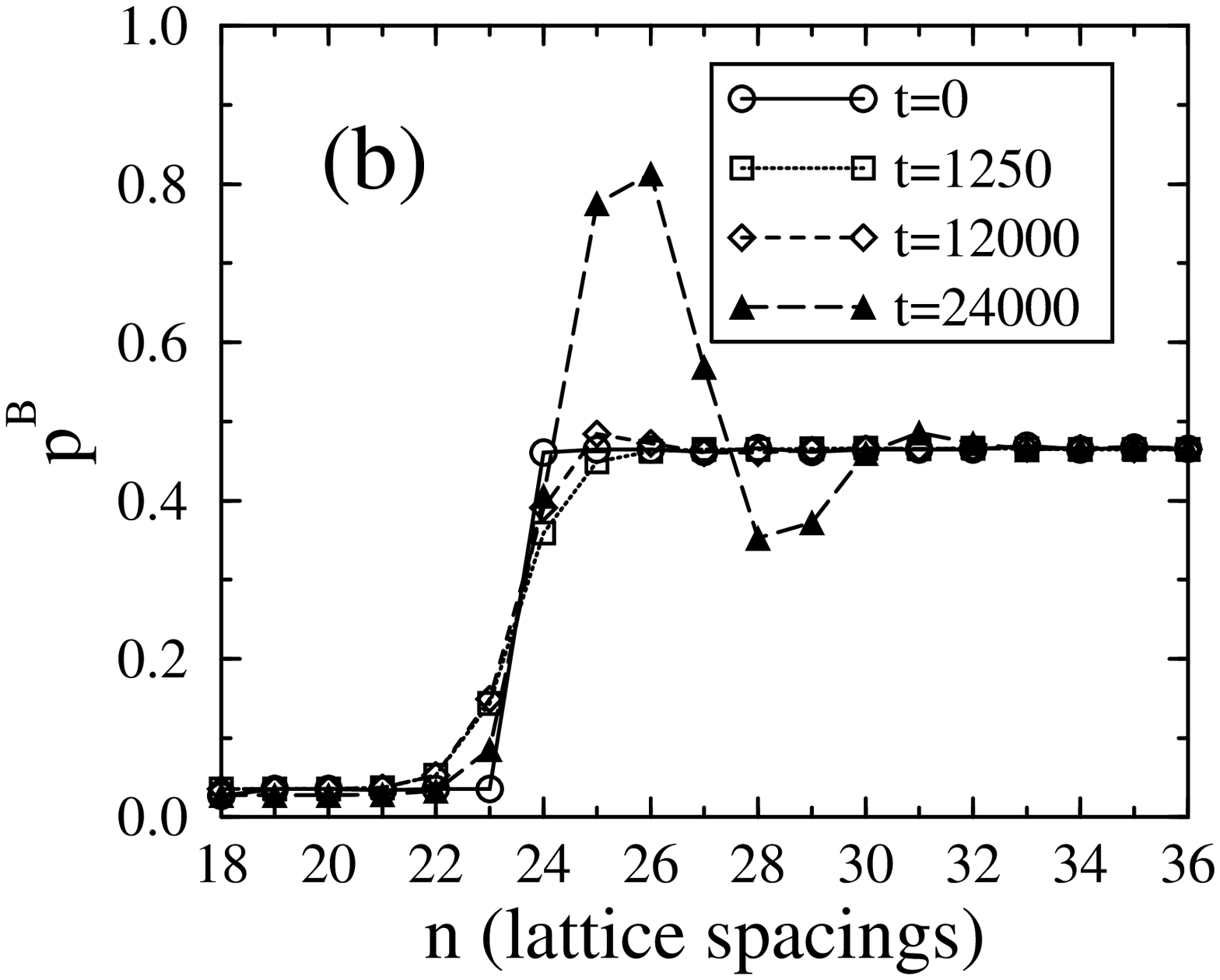,width=0.4\textwidth}
 \hskip .5in}
\caption{Plots of $p^A_n$ (a) and $p^B_n$ (b) along a line normal to
a flat interface. From the initial step profile, concentrations
rapidly relax to a smooth profile. Then, on a longer time scale,
phase separation of A and B occurs, starting from the surface. 
Times in units of $w_0^{-1}$.}
\label{figintersym}
\end{figure}

To see more in detail what happens at the interface, we plot in
Fig. \ref{figintersym} several snapshots of the concentration
profiles along a line which is normal to the surface at a 
randomly chosen point. The initial step profile quickly relaxes
to a smooth shape and stays nearly stationary, until 
on a slower time scale B is enriched and
A depleted; at other interface points, the opposite happens. 
An oscillatory concentration profile develops, with an
amplitude which has its maximum in the interface and decays
into the solid. We will show below that the envelope of this
oscillation becomes a decaying  exponential away from 
the surface. The fact that the perturbation decays
exponentially with the distance from the surface, but
grows exponentially with time, explains the constant
propagation velocity of the decomposition front.
Note that the concentrations in the
vapor vary only very slightly: the vapor is a stable phase,
and hence perturbations decay. Because of the fast diffusion
in the vapor, the surroundings of the droplet act as a
particle reservoir.

\begin{figure}
\centerline{
 \psfig{file=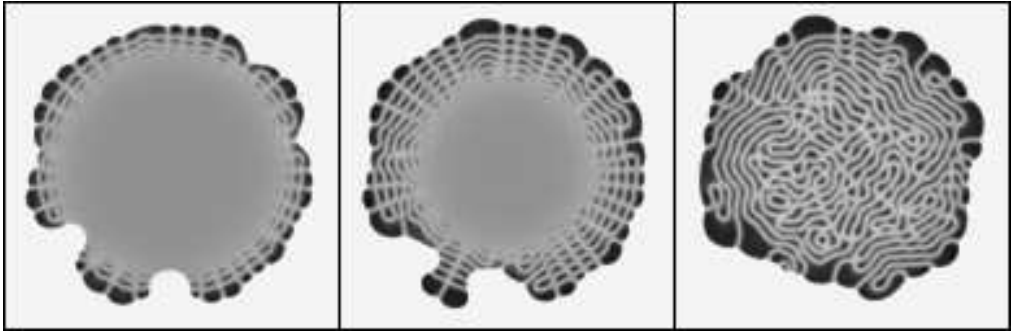,width=0.75\textwidth}}
\medskip
\caption{Vacancy distribution during the simulation of Fig.
\ref{figdroplet}. Dark areas have low vacancy content. The
contrast inside the droplet has been enhanced. Vacancies
accumulate in the domain boundaries. The pictures correspond
to snapshots number 3,4, and 6 in Fig. \ref{figdroplet}
($t = 49400$, $74400$, and $149200$ $w_0^{-1}$).}
\label{figvac}
\end{figure}

As the surface mode develops, the mixture-vapor interface
deforms, and fingers of vapor start to grow into the
interior of the droplet. This is the result of a Mullins-Sekerka
instability \cite{Mullins64} with respect to the vacancies.
To clarify this point, we show in Fig. \ref{figvac} snapshots
of the vacancy concentration during the decomposition process.
At the beginning, the vacancy concentration
stays constant and equal to the initial value. Once the decomposition
process reaches its nonlinear stage, however, vacancies are
expelled from the domains of the new equilibrium phases.
These excess vacancies have to diffuse to the surface of
the droplet.
A finger of vapor protruding into the mixture 
enhances the concentration gradients around its tip, 
and hence grows faster than a flat portion of the surface.
In addition, the diffusion is faster in the initial
mixture than in the phase-separated domains. The
growth of the fingers stops when
a layer of decomposed material has formed around their
entire contour. The diffusion then takes place mainly along 
the domain boundaries, where the vacancies are enriched.
The fingers are smoothed out by the subsequent coarsening process. 

The propagation of the surface mode stops when the bulk modes enter 
the nonlinear regime. The structures in the interior 
of the sample are the usual bicontinuous patterns of bulk spinodal
decomposition at equal volume fractions. Both surface and bulk
structures coarsen by the evaporation-condensation
mechanism. This process is faster at the surface because of the
rapid diffusion through the vapor. At the exterior 
surface of the droplet, there exist trijunction
points where the three phases are in contact. The angles between 
the interfaces at these points are fixed by the local 
equilibrium between the three surface tensions of Av, Bv, and 
AB-interfaces. For our symmetric choice
of interaction energies, all angles are $120^\circ$ in local equilibrium.

The initial values for the concentrations in our example represent
a special choice: the chemical potentials of the two
species and the grand potential have the 
same value in the mixture and in 
the vapor. There exists exactly one set of concentrations
satisfying these conditions at a given temperature. In this
special case, there is no net mass flux between vapor and mixture,
and the interface is at rest. This choice was mainly made to
simplify the stability calculations to be presented below.
The existence of a surface mode, however, is not limited to
this special case. For other initial conditions (but still
$\pasol=\pbsol$ and $\pavap=\pbvap$), a surface mode
develops while the mixture-vapor interface slowly moves. 

\begin{figure}
\centerline{
 \psfig{file=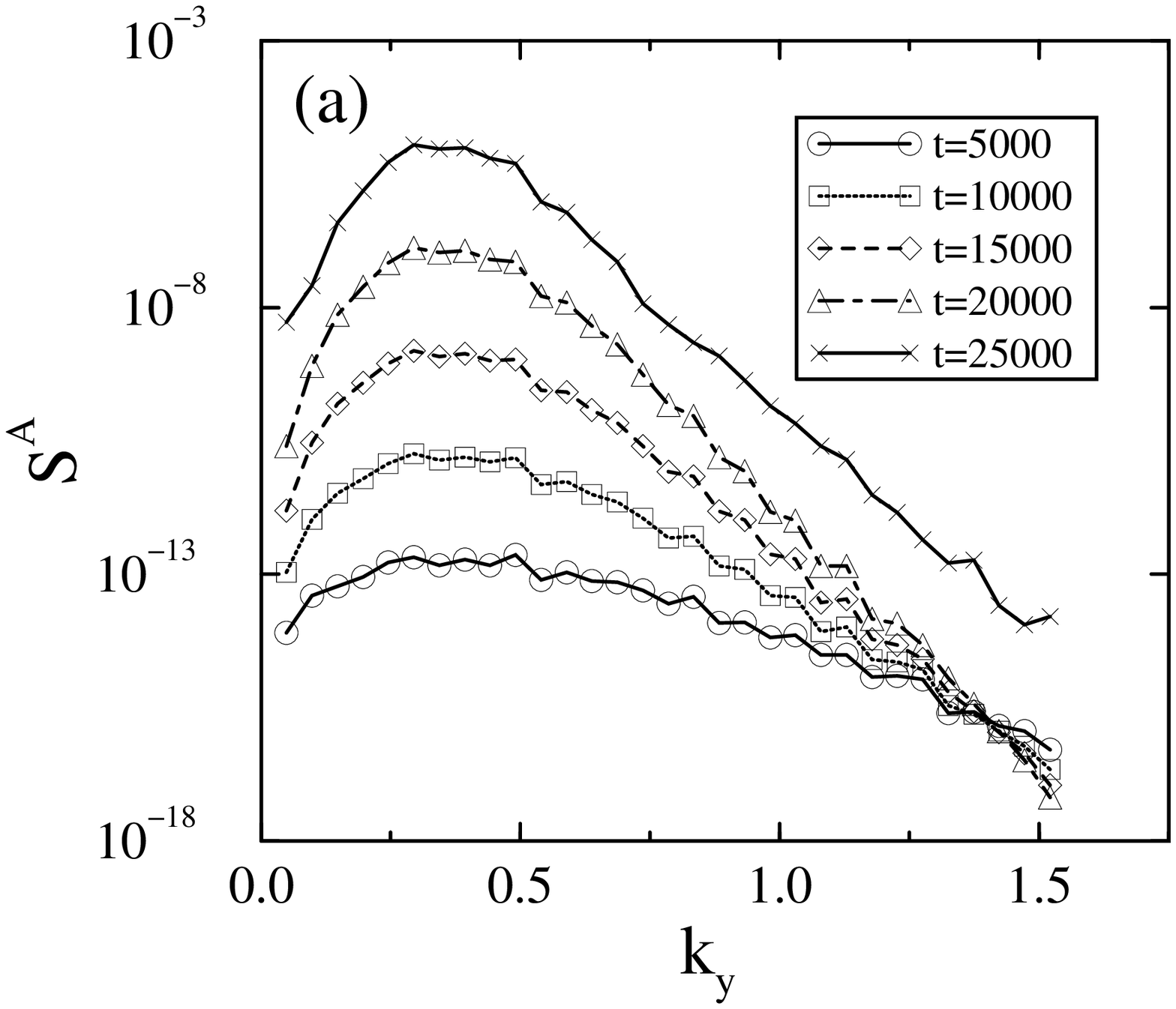,width=0.45\textwidth}\hfill
 \psfig{file=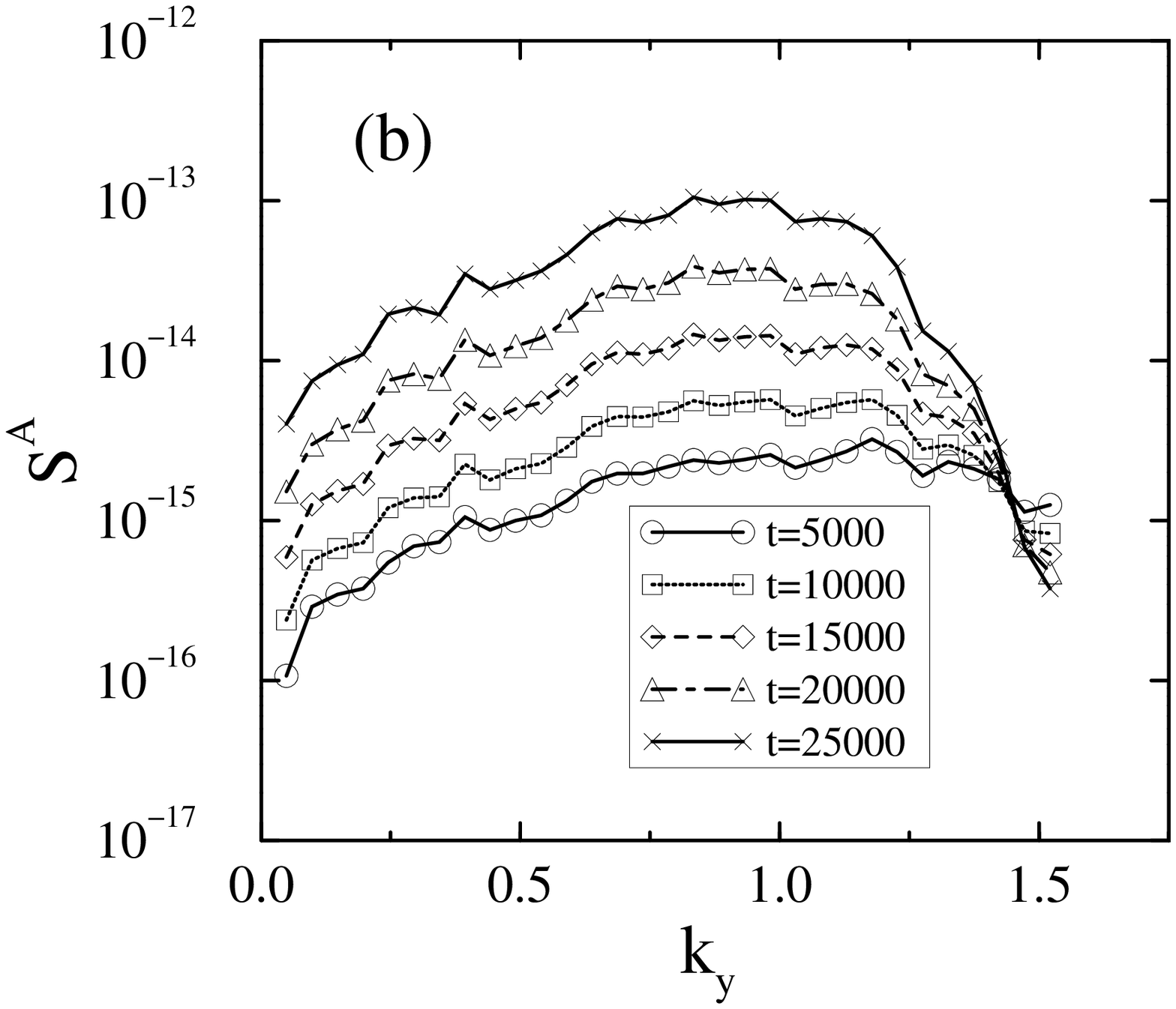,width=0.45\textwidth}}
\caption{Plots of $S^A(k_y,x,t)$ (a) for the first solid
layer and (b) in the middle of the slab. The data are
averaged over 10 runs with different initial fluctuations 
(noise amplitude $A_0 = 0.5\times 10^{-5}$).}
\label{figstruct}
\end{figure}

To obtain more quantitative information about the phase 
separation process, we repeated our simulation with the same 
parameters, but this time in a stripe geometry: a slab of mixture
along the $y$-direction is immersed in the vapor; we thus have
two straight interfaces normal to the $x$ axis. 
A convenient quantity for the analysis of phase separation processes
is the dynamical structure factor. Of particular interest in our case
is the difference between bulk and surface behavior. Therefore, we
define a one-dimensional structure factor along the interface:
\begeq
S^\alpha(k_y,x,t)=
    \left|{1\over L_y}\sum_{\{j | x_j=x\}} p_j^\alpha e^{ik_y y_j}\right|^2,
\endeq
where $x_j$ and $y_j$ are the coordinates of lattice site $j$. The
sum goes over a lattice plane (in 2D: a line) at a fixed $x$-coordinate,
and $k_y$ is parallel to the surface. Figure \ref{figstruct} shows
plots of this quantity for different times.
We observe the characteristic
Cahn-Hilliard behavior: in the beginning, linear superposition
is valid, and the structure factor for each mode
grows exponentially with a growth rate depending on $k_y$.
In particular, all the structure factor curves intersect in one
point, corresponding to the marginally stable mode.
At later times, the nonlinear terms in the equations of 
motion couple the different modes, and the shape of the
structure factor curve changes, as can be noted at the
last time for the surface layer. The two sets of curves
have very different amplitudes, and the
maximum is located at $ak_y\approx 0.3$ at the
surface, and at $ak_y\approx 1$ in the bulk.
These differences between bulk and
surface behavior will be addressed in Sec. 4.

\begin{figure}
\centerline{
 \psfig{file=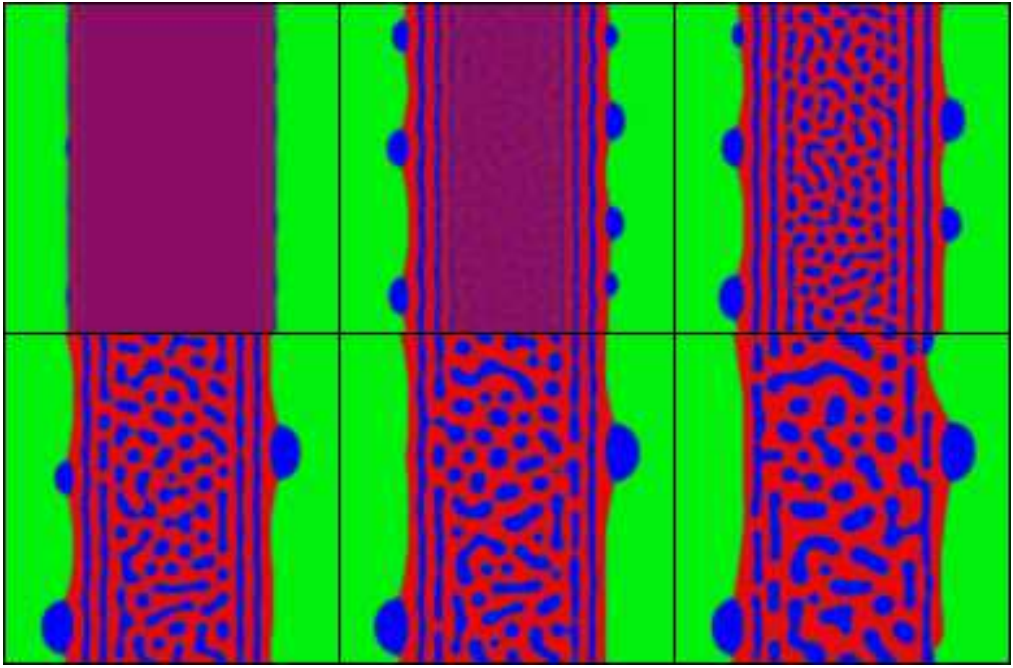,width=0.75\textwidth}}
\medskip
\caption{Snapshot pictures of a mixture with $\pasol = 0.56007$,
$\pbsol = 0.37338$ in a vapor of  $\pavap = 0.04507$, $\pbvap = 0.03019$.
The other parameters and the colors are as in Fig. \ref{figdroplet}.
$t = 10600$, $60700$, $123200$, $248300$, $373400$, and $623500$ $w_0^{-1}$.}
\label{figstripe}
\end{figure}

Very different structures appear when the concentration 
of the mixture is sufficiently off-critical. 
Fig. \ref{figstripe} shows the evolution of a 
slab of an AB mixture with a concentration ratio 60:40. 
The initial concentrations were again chosen to give equal 
chemical potentials and grand potential in the two bulk phases; 
note that now $\mu^A\neq\mu^B$.
The minority component rapidly segregates
at the surface, triggering a ``spinodal wave'' 
normal to the surface. A surface mode with modulations
along the interface is still present and leads to a
destabilization of the first layer of the minority 
component, which disintegrates into regularly spaced droplets. 
These droplets coarsen rapidly. This time, there
is no Mullins-Sekerka instability with respect to the
vacancies, because the formation of the first decomposed
layer rapidly blocks the exchange of vacancies between
the interior of the sample and the vapor.

Inside the sample, the surface-induced wave travels until 
the bulk modes reach their nonlinear regime. There is a
competition of droplet and stripe patterns during the
subsequent coarsening process. The droplets in the 
interior of the sample coarsen more rapidly than 
the stripes at the surface, not surprisingly as the
driving force for the evaporation-condensation mechanism
of coarsening is the curvature of the domain walls.
Ultimately, the stripes start to break up
and are ``infected'' by the droplet pattern.

\begin{figure}
\centerline{
 \psfig{file=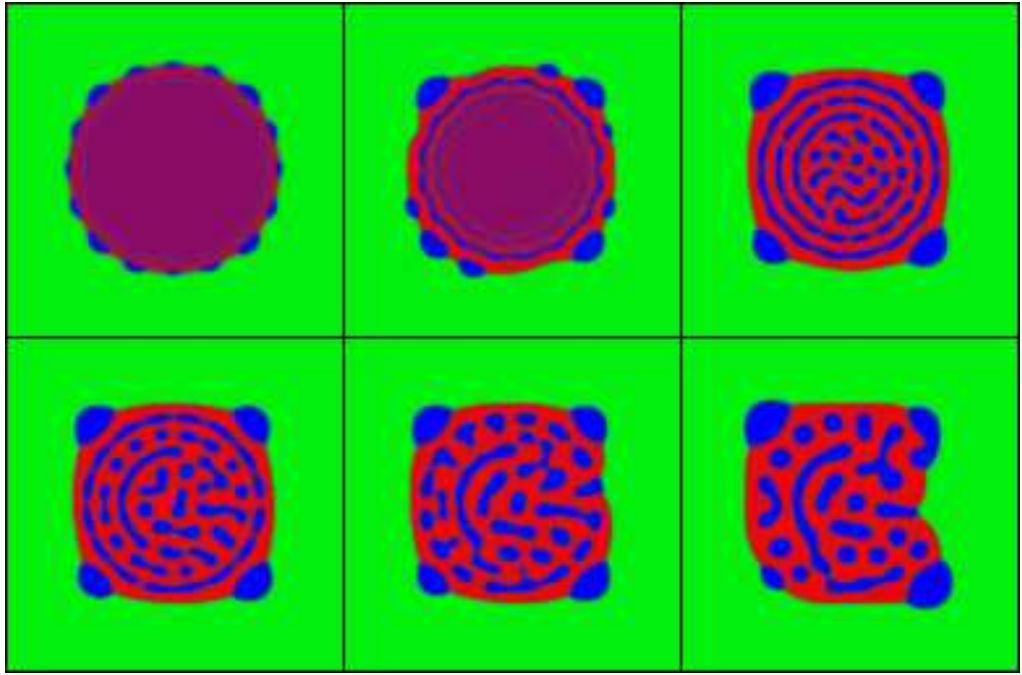,width=0.75\textwidth}}
\medskip
\caption{Same parameters as the preceding figure, but in circular
geometry. The symmetries in the picture are a consequence of the
lattice structure. Times: $10600$, $35600$, 
$123200$, $248300$, $373400$, and $623500$ $w_0^{-1}$}
\label{figtarget}
\end{figure}

A simulation for a ``droplet'' geometry is shown in 
Fig. \ref{figtarget}. The obviously symmetric configuration
of the outer domains of B-rich phase is due to the anisotropy
of the surface tensions introduced by the lattice \cite{Plapp97a}.
It is interesting to note that this effect is not
immediately visible in the critical quenches. The
symmetric configuration, however, is only transient: on the last snapshot
picture, the smallest of the four outer B domains is about to evaporate.
Note also the symmetry in the first and second ring of inner droplets; 
this pattern is later destroyed by the coarsening of the bulk structures.

\begin{figure}
\centerline{
 \hskip .5in
 \psfig{file=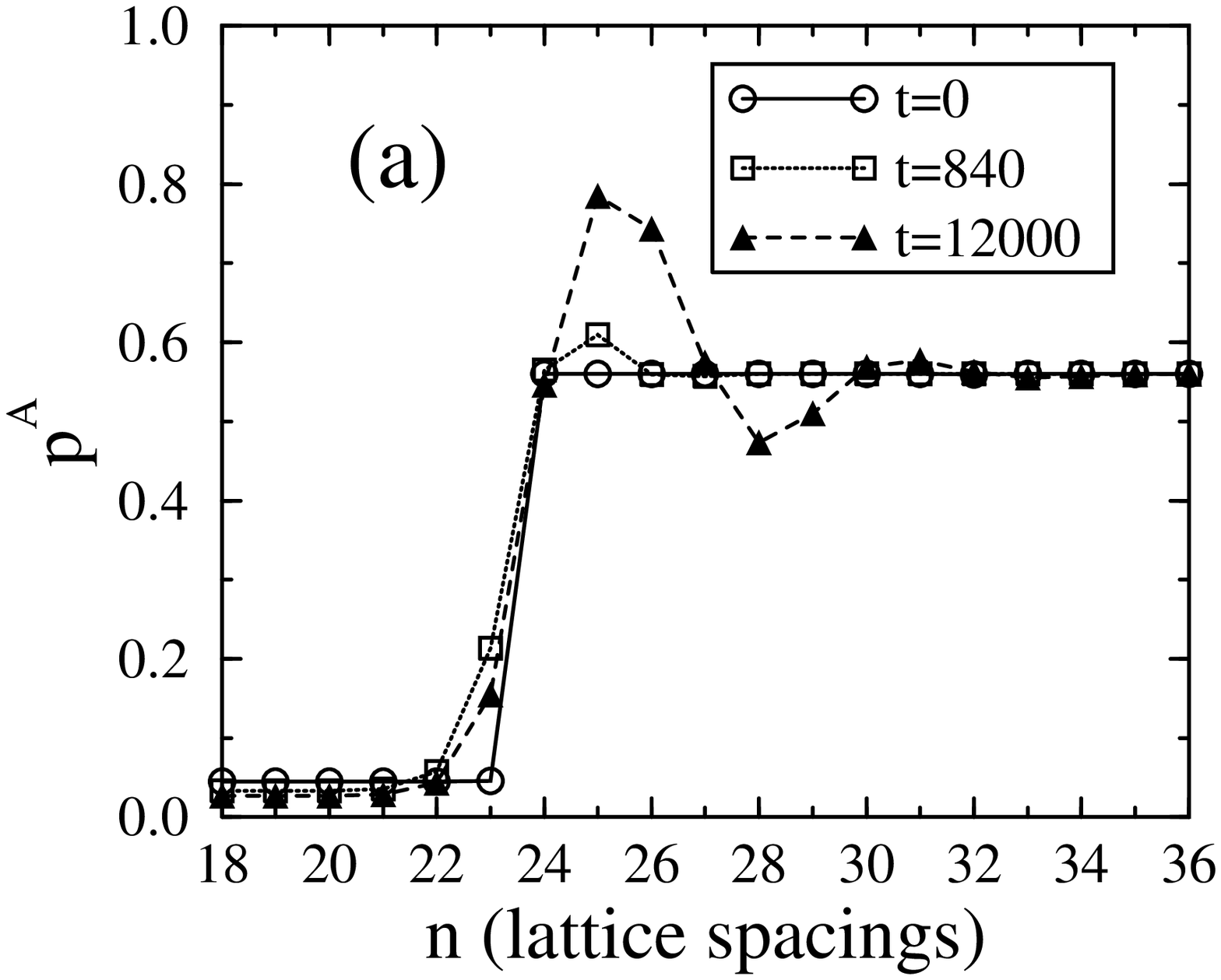,width=0.4\textwidth}\hfill
 \psfig{file=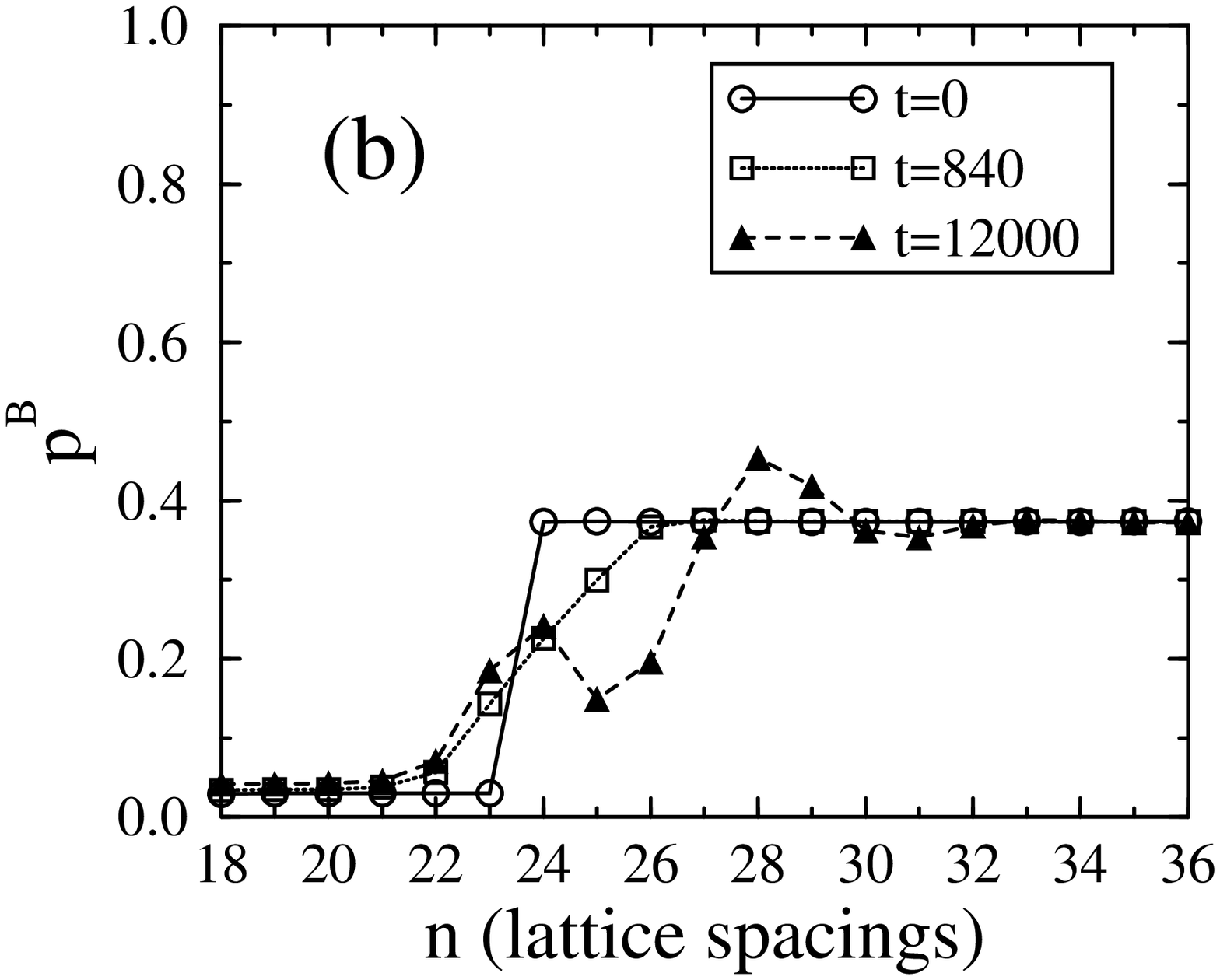,width=0.4\textwidth}
 \hskip .5in}
\caption{Same as Figure \ref{figintersym}, but for an off-critical
quench. Note the faster evolution. Times in units of $w_0^{-1}$.}
\label{figinterasym}
\end{figure}

Figure \ref{figinterasym} shows the evolution of the interface
profiles. In contrast to the critical quench, there is no smooth,
nearly stationary state at intermediate times. The profile
immediately starts to show the onset of oscillations. Also, the 
evolution is faster: at $t=12000 w_0^{-1}$, when the separation of 
A and B is still tiny in Fig. \ref{figintersym}, it is already well
pronounced in the off-critical case.

\begin{figure}
\centerline{
 \psfig{file=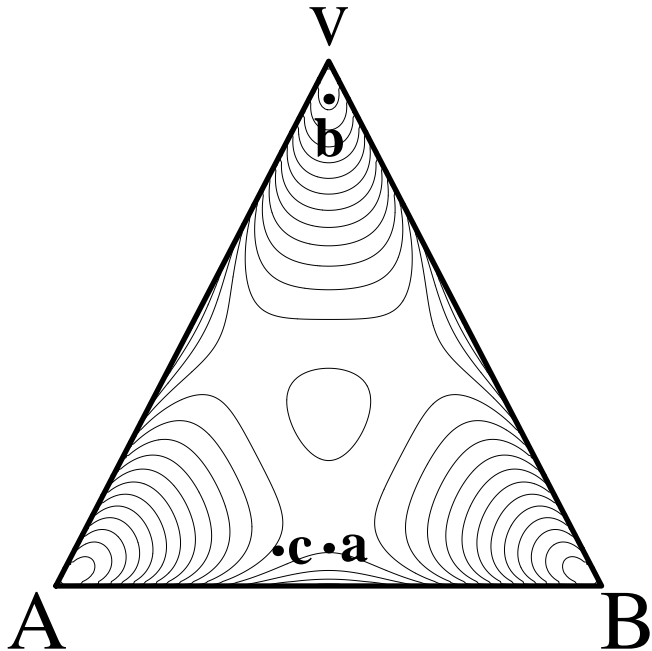,width=0.5\textwidth}}
\caption{Free energy surface, represented by its isoenergy
lines, for $kT/\epsaa=0.5$. The three minima near the corners
are separated by a ``hill'' in the center.}
\label{figfsurf}
\end{figure}

The difference between the two evolutions can be qualitatively 
understood by a look on the free energy surface, plotted in
Fig. \ref{figfsurf}. In this figure, the points a, b, and c
denote the initial compositions of the critical and off-critical
mixtures and the vapor, respectively. In the symmetric case,
we can connect the points a and b along a symmetry axis of the
free energy density, with zero slope along the A-B-direction.
This means that an interface which is situated
completely on this line is stationary with 
respect to a separation of A and B.
For the case of an off-critical concentration, we cannot find
any such line going from b to c, and an unstable stationary
interface does not exist. At the surface of the mixture,
chemical potential gradients will always lead to a segregation
of one of the components to the surface.
Which component is attracted to the surface 
depends on the choice of concentrations 
in the vapor phase. Let us mention that this argument
is not entirely complete, because it considers only the
free energy density, whereas the complete free energy also
contains the discrete gradient terms. In our simulations,
however, we never observed any unstable stationary interface
configuration for off-critical compositions (or asymmetric
interaction energies).

The transition from the checkerboard structures to the stripes is
gradual. For slightly asymmetric concentrations, the segregation to the
surface is slow, and the surface mode has enough time to grow. We
observed checkerboard structures up to a concentration ratio of
approximately 54:46. For this composition,
checkerboard structures and stripes appear simultaneously
on different portions of the surface. Similar findings are also valid
when we vary the interaction parameters in our model: for
$\epsbb \neq \epsaa$, we have always observed surface-directed
spinodal waves; however, a surface mode should appear in 
this case if for some asymmetric compositions the surface
segregation becomes slow.

These findings are consistent with calculations for mixtures near flat
substrates using continuous equations of Cahn-Hilliard type: spinodal
waves occur when one component is attracted to the substrate 
\cite{Puri94,Frisch95}, whereas surface modes have been found in the
case of a substrate which prefers neither of the components of the
mixture \cite{Fischer97}. Our simple model shows that this
general behavior is also valid for free surfaces.

\begin{figure}
\centerline{
 \psfig{file=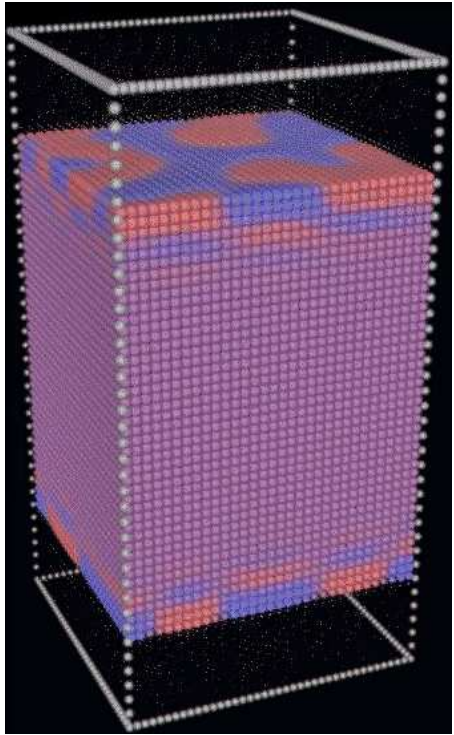,width=0.33\textwidth}}
\medskip
\caption{3D sample on a $32\times 32\times 64$ lattice
at critical composition (initial concentrations as in 
Fig. \ref{figdroplet}, $kT/\epsaa=0.75$). The outer white spheres 
indicate the limits of the simulation box. Periodic boundary conditions
are applied in all directions. The size of each sphere inside 
the box indicates the total concentration ($p^A + p^B$), 
the color its composition (red: A, blue: B).
Regions of vapor are situated above and below the mixture
film; as the concentration of atoms is very low in these
regions, the corresponding spheres are tiny.
This snapshot was taken at $t=20400 w_0^{-1}$.}
\label{figtdsym}
\end{figure}

\begin{figure}
\centerline{
 \psfig{file=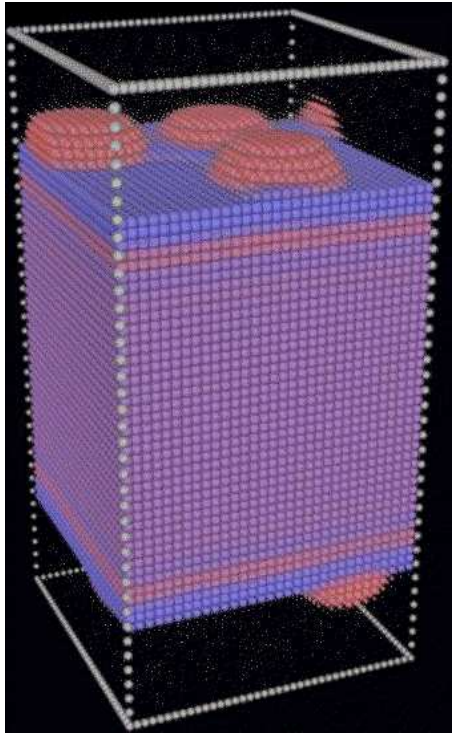,width=0.33\textwidth}}
\medskip
\caption{3D sample at off-critical
composition. Initial concentrations are as in 
Fig. \ref{figstripe}, $kT/\epsaa=0.75$, and 
$t=20400 w_0^{-1}$.}
\label{figtdasym}
\end{figure}

In 3D samples we also find fast surface modes, 
but because the vapor-mixture interface is now
two-dimensional, the patterns occurring at the surface
for critical quenches are
those of 2D bulk spinodal decomposition (Fig. \ref{figtdsym}). A
bicontinuous pattern forms at the surface and propagates into the
sample, replicating itself in an oscillatory manner, that is we always
find an oscillating concentration profile when we look normal to the
surface. In the off-critical case, a spinodal wave occurs as
in two dimensions (Fig. \ref{figtdasym}).

\begin{figure}
\centerline{
 \psfig{file=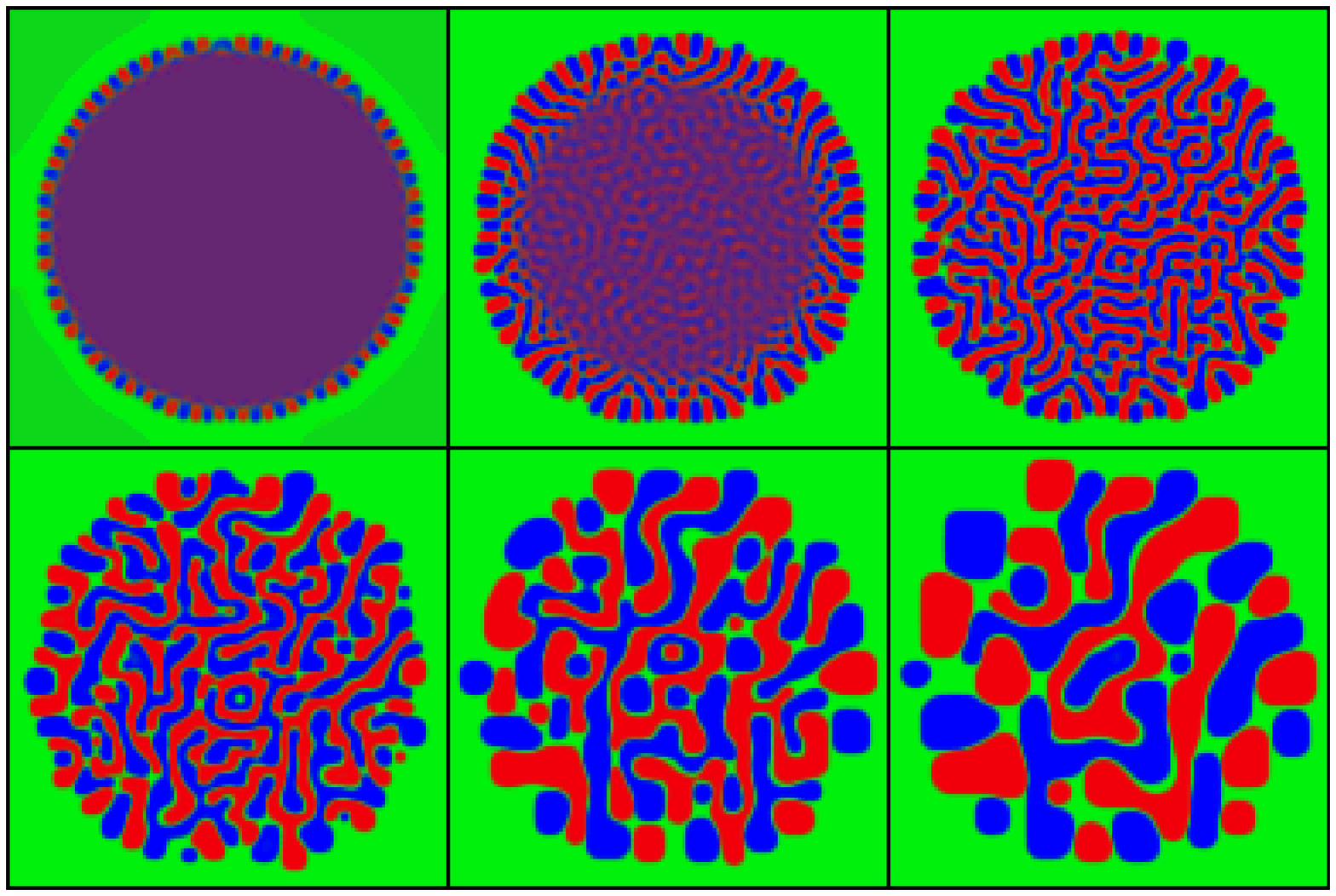,width=0.75\textwidth}}
\medskip
\caption{Decomposition of a droplet of initial compositions
$\pasol=\pbsol=0.4276$ in a vapor with $\pavap=\pbvap=0.0724$
for $\epsaa=\epsbb=1, \epsbb=0, kT=0.4$. The snapshot times
are $t = 560$, $1590$, $4130$, $14100$, $48830$, and
$149010$ $w_0^{-1}$.}
\label{figeclat}
\end{figure}

In all the simulations presented so far, the domains 
of the A- and B-rich phases stick together, because 
the surface tension of an Av- or Bv-interface
is more than the half of the one of an AB-interface. 
This changes if we lower the interaction energy $\epsab$: 
for $\epsab = 0$, the droplet ``explodes'': thin layers 
of vapor penetrate into the interior of the droplet 
along the forming AB-interfaces, and the domains of A and
B are slowly drifting apart (Fig. \ref{figeclat}). 
This reminds of the decomposition
of a binary mixture in the presence of a surfactant.
The presence of sharp corners and facets in the
domain shapes of the last snapshot indicates that the 
surface tension anisotropy is quite large.

\section{Linear stability analysis}
\nobreak
\subsection{Bulk}
\nobreak
\noindent
We will now study more in detail the checkerboard structures. To this
end, we must calculate the growth rates of bulk and surface modes. We
start with the bulk modes. A homogeneous system of overall
composition $\bar p^A$ and $\bar p^B$ is perturbed by small
fluctuations of the occupation probabilities:
\begeq
p_i^\alpha(t) = \bar p^\alpha + \delta_i^\alpha(t) \quad (\alpha = A,B),
\endeq
with $\delta_i^\alpha \ll 1$. We linearize the chemical potentials
around the average concentrations. Introducing a vector notation
with respect to the two species of particles, we obtain starting
from Eqs. (\ref{chempots})
\begeq
\begvec \mu_i^A \\ \mu_i^B \endvec = \begvec \bar\mu^A \\ \bar\mu^B \endvec
      - {\bf E}
      \begvec \lap\delta_i^A \\ \lap\delta_i^B \endvec + 
      {\bf S} \begvec \delta_i^A \\ \delta_i^B \endvec.
\endeq
Here, $\bar\mu^\alpha$ are the unperturbed values of the 
chemical potentials, the matrix ${\bf E}$ of the interaction energies is
\begeq
{\bf E} = \begmat \epsaa & \epsab \\ \epsab & \epsbb \endmat,
\endeq
and $S$ is the matrix of the second derivatives
of the free energy density, taken at $\bar p^A$ and $\bar p^B$:
\begeq
{\bf S} = \begmat S_{AA} & S_{AB} \\ S_{AB} & S_{BB}\endmat
\quad {\rm with} \quad
S_{\alpha\beta} = {\left.{\partial^2 f\over 
     \partial p^\alpha \partial p^\beta}\right|}_{\bar p^A, \bar p^B}.
\endeq
The variations in the chemical potentials create currents. To
obtain these currents to order one in $\delta_i^\alpha$, we may
use the unperturbed values of the mobilities,
\begeq
M_{ij}^\alpha = M_{\rm hom}^\alpha(\bar p^A,\bar p^B) = \bar M^\alpha.
\endeq
The equations of motion become:
\begeq
{d\over dt} \begvec \delta_i^A \\ \delta_i^B \endvec = 
\begvec \bar M^A \lap \mu_i^A \\ \bar M^B \lap \mu_i^B \endvec.
\endeq
As a homogeneous system is translation invariant 
with respect to the lattice vectors,
solutions of the linearized equations are of the form:
\begeq
\begvec \delta_j^A \\ \delta_j^B \endvec = \begvec \delta^A \\ \delta^B \endvec
    \exp\left(i\vec k \cdot \vec x_j + \omk t\right),
\endeq
where $\vec x_j$ is the position vector of site $j$ in real space,
and $\vec k = (k_x,k_y)$ is the wave vector of the perturbation.
The growth rate $\omk$ and the coefficients $\delta^\alpha$ have to
be determined by solving the eigenvalue problem
\begeq
\omk \begvec \delta^A \\ \delta^B \endvec =
\Ak \begmat  
   \bar M^A & 0 \\ 0 & \bar M^B \endmat \,
   \left({\bf S} - \Ak{\bf E}\right)
   \begvec \delta^A \\ \delta^B \endvec.
\label{disprel}
\endeq
Here, the terms 
\begeq
\Ak = -4\sin^2(k_xa/2) -4\sin^2(k_ya/2)
\label{ak}
\endeq
arise from the discrete Laplacians. Eq. (\ref{disprel}) is
quadratic in $\omega$ and thus gives a stability spectrum
with two branches. Each of these branches can be stable
($\omega$ is negative for all wave vectors) or unstable.
In the latter case, positive growth rates occur for small
values of $|\vec k|$. The number of unstable branches is
equal to the number of negative eigenvalues of the matrix
$\bf S$. This can be easily seen by taking the limit 
$|\vec k|\to 0$ in Eq. (\ref{disprel}). The terms proportional
to $\Ak^2$ can be neglected, and we have
\begeq
\omk \begvec \delta^A \\ \delta^B \endvec =
   -{|\vec k|}^2 {\bf S}\begvec \delta^A \\ \delta^B \endvec.
\endeq
The matrix {\bf S} is related to the curvature of the free energy
surface. If both eigenvalues are positive, the surface is
locally convex, and the homogeneous state is stable. For
two eigenvalues of different sign, the surface has locally
the structure of a saddle point, and we have partial
instability: only fluctuations in the
concave direction in concentration space are amplified. 
Finally, for two negative eigenvalues, the free energy
surface is concave and all perturbations grow. The frontiers
between these regions of different stability behavior are
given by the spinodal surfaces, defined by the condition
\begeq
\det{\bf S} = 0.
\endeq
This generalizes the concept of a spinodal curve to our
three-component system.

\begin{figure}
\centerline{
 \psfig{file=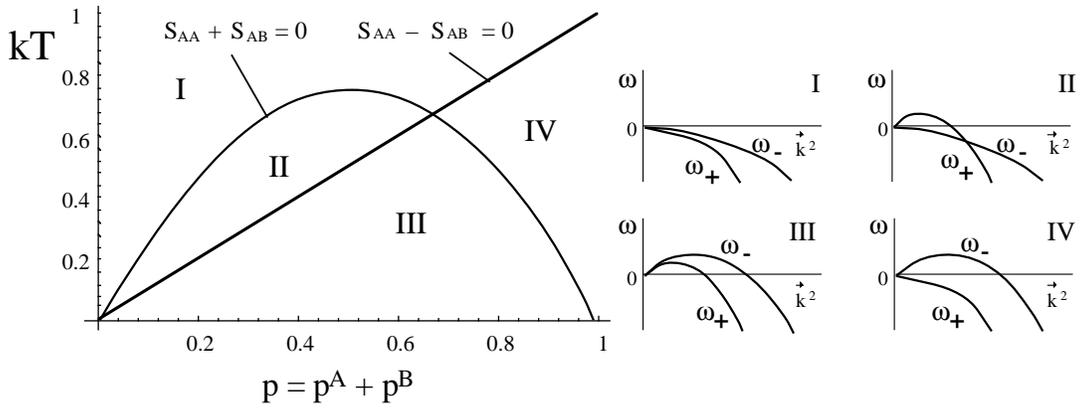,width=0.8\textwidth}}
\caption{Cut through the spinodal surfaces along the plane
$p^A = p^B$ (left) and the typical behavior of the dispersion
relations $\omega_+(\vec k)$ and $\omega_-(\vec k)$ (right). 
In region I, the system is stable, in region II it is unstable
against separation of particles and vacancies, in region IV
against separation of A and B; in region III, it is completely
unstable.}
\label{figspdiag}
\end{figure}
For the special case of symmetric interaction energies and
equal average concentration, Eq. (\ref{disprel}) can be
considerably simplified because then we have 
$\bar M^A = \bar M^B = \bar M$. In addition, the matrix $\bf L$ 
defined by
\begeq
{\bf L} = \Ak ( {\bf S} - \Ak {\bf E})
\endeq
is symmetric. We obtain immediately
\begeq
\omega_{\pm}(\vec k) = \bar M (L_{AA} \pm L_{AB}),
\label{omksym}
\endeq
and the associated eigenvectors are
\begeq
\begvec \delta^A \\ \delta^B \endvec = \begvec 1 \\ \pm1 \endvec.
\endeq
This last result shows that of the two branches of the dispersion
relation, the first describes the separation into a dense AB mixture
and a dilute vapor, whereas the second gives a separation between
A and B, leaving the local vacancy concentration unchanged.
Fig. \ref{figspdiag} shows a cut through the spinodal surfaces along
the axis $\bar p^A = \bar p^B$. We also indicate the regions of different
stability behaviors, and show the typical shapes of $\omega_+(\vec k)$
and $\omega_-(\vec k)$ in these regions.
The interior of our samples is in the region where only
the mode separating A and B is unstable: the local vacancy 
concentration stays unchanged in the linear stage. Other modes of
decomposition in a homogeneous ternary system have been studied
by Chen \cite{Chen94b}.

The anisotropy due to the lattice structure enters in the above
formulae by the factor $\Ak$. For the long wavelength perturbations
considered here ($a|\vec k| < 1$), the relative variations of $\Ak$ with
orientation are of the order of a percent. We will therefore
neglect this dependence and use a wave vector along one of the lattice
directions for comparisons to numerical results. For the example
of the simulation shown in Fig. \ref{figdroplet}, i.e. with
$kT/\epsaa = 0.5$ and $\bar p^A = \bar p^B = 0.46464$, we found
a maximum bulk growth rate of $\omega_b = 1.06\times 10^{-4} w_0$
at a wave number $k_b = 1.003/a$.

\subsection{Surface}
\nobreak
\noindent
We must now analyze the stability spectrum at the surface. This
is more difficult than in the bulk because the initial state is
now heterogeneous. To simplify the problem, we will treat the
case of a flat interface which is normal to one of the lattice
directions, say $x$. Then, in the initial state all lattice sites
in a layer at a given $x$ coordinate have the same
concentrations. In what follows, we will replace the site indices
``$i$'' used so far by a pair of indices ``$n,j$'', where $n$
numbers the layer, and $j$ numbers the sites in the $y$ direction.
We use the same values of the concentrations in the bulk phases
as for our simulations. As the chemical potentials
are equal in the two phases, the initial interface state can be obtained
by fixing the chemical potentials to their appropriate values and
numerically solving the one-dimensional version of the finite
difference equations Eq. (\ref{chempots}) for the concentrations
in the $n$th layer $\bar p^\alpha_n$.

Whereas the translation invariance is broken along the $x$-axis,
it is preserved along $y$, and hence we can still use a Fourier
representation. We write the perturbed state as
\begeq
p^\alpha_{n,j}(t) = \bar p^\alpha_n + \delta_{n,j}^\alpha(t),
\endeq
\begeq
\delta_{n,j}^\alpha(t) = \delta_n^\alpha \exp\left(ik_yja + \omky t\right),
\label{fouriery}
\endeq
with two unknown coefficients $\delta^A_n$ and $\delta^B_n$ per layer.
We start with the linearization of the chemical potentials. The discrete
Laplacians appearing in Eq. (\ref{chempots}) give contributions of
the form
\begarr
\lap\delta_{n,j}^\alpha & = & 
    \delta_{n+1,j}^\alpha + \delta_{n-1,j}^\alpha + 
    \delta_{n,j+1}^\alpha + \delta_{n,j-1}^\alpha - 
    4\delta_{n,j}^\alpha \nonumber \\
    & = & \left[\delta^\alpha_{n-1} + \delta^\alpha_{n+1} + 
          \left(\Ak-2\right)\delta_n^\alpha\right] \,
               \exp\left(ik_yja + \omky t\right).
\endarr
Denoting by $\bar\mu^\alpha$ the (constant) values of the unperturbed
chemical potential, and by ${\bf S}(n)$ the matrix of the second
derivatives, taken at $\bar p_n^A$ and $\bar p_n^B$, we find
\begeq
\begvec \mu^A_{n,j}-\bar\mu^A \\ \mu^B_{n,j}-\bar\mu^B \endvec = 
  \left({\bf S}(n) - (\Ak - 2) {\bf E}\right) 
     \begvec \delta^A_{n,j} \\ \delta^B_{n,j} \endvec
  - {\bf E} \begvec \delta^A_{n-1,j} \\ \delta^B_{n-1,j} \endvec
  - {\bf E} \begvec \delta^A_{n+1,j} \\ \delta^B_{n-1,j} \endvec.
\label{mumulin}
\endeq
The next step is the linearization of the equations of motion.
As in the homogeneous case, the mobilities can be taken in the
unperturbed initial state. But because the concentrations vary
through the interface, we cannot use the expression for the
homogeneous system. The appropriate expressions can be obtained
from Eq. (\ref{mobil}) by taking the limit $\mu_j\to\mu_i$ with
$p_j\neq p_i$:
\begeq
\bar M_{mn}^\alpha = w_0 
    {(1-\bar p_m^A-\bar p_m^B)(1-\bar p_n^A-\bar p_n^B)\over kT}
     \exp\left(\bar \mu^\alpha\over kT\right)
\endeq
(the two indices of $\bar M_{nm}$ are two layer indices).
The linearized equation of motion becomes
\begeq
{d \over dt} \delta_{n,j}^\alpha =
         \bar M^\alpha_{nn}\left(\mu^\alpha_{n,j+1}-2\mu^\alpha_{n,j}+
             \mu^\alpha_{n,j-1}\right)
      +\bar M^\alpha_{nn+1}\left(
       \mu^\alpha_{n+1,j}-\mu^\alpha_{n,j}\right)
      +\bar M^\alpha_{nn-1}\left(
       \mu^\alpha_{n-1,j}-\mu^\alpha_{n,j}\right).
\endeq
Inserting Eqs. (\ref{mumulin}) and (\ref{fouriery}), one obtains
an eigenvalue problem for $\omega$, the eigenvectors given by the
set of unknown coefficients $\delta_n^\alpha$. This means that
the matrix to diagonalize is of size $2L\times 2L$, where $L$ is
the number of layers in the $x$ direction, and that the corresponding
dispersion relation has $2L$ branches. This is a standard linear
algebra problem and can be solved numerically.

\begin{figure}
\centerline{
 \psfig{file=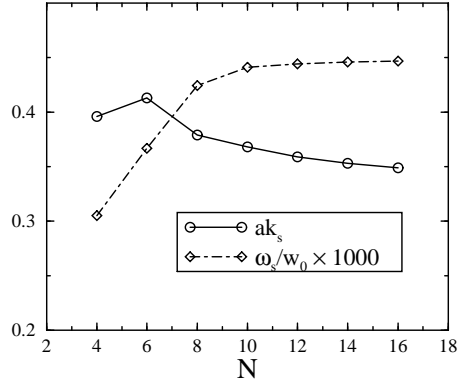,width=0.4\textwidth}}
\caption{Maximum growth rate $\omega_s$ and corresponding wave
vector $k_s$ for the vapor-mixture interface with initial
conditions as in Fig. \ref{figdroplet} as a function of
the number $N$ of layers around the interface included in the
eigenvalue problem.}
\label{fignumtest}
\end{figure}
In the initial stage of the development, the surface mode is
localized in a small number of layers around the surface. Hence we 
can simplify the problem by considering a small ``solution region''
around the interface. We assume perturbations outside a narrow
region of $N$ layers centered around the interface to be zero.
We diagonalized the resulting $2N\times 2N$ matrix numerically 
and obtained the maximum growth rate, $\omega_s$, and the 
corresponding wave vector $k_s$. As shown in Fig. \ref{fignumtest},
these quantities become independent of $N$ if a sufficient number
of layers is included \cite{Note}.

\begin{figure}
\centerline{
 \psfig{file=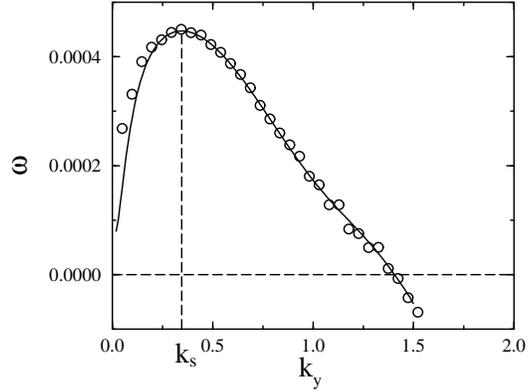,width=0.4\textwidth}}
\caption{Comparison of theoretical prediction (line) and simulation 
results (circles) for the stability spectrum of the vapor-mixture 
interface with the same parameters 
as in Fig. \ref{figdroplet}. The theoretical curve
is calculated with $N=16$, the growth rates in the simulation
were extracted from the data in Fig. \ref{figstruct}. The scatter
in the simulation data is due to the finite number of different
runs.}
\label{figcomp}
\end{figure}
To compare the results of this stability analysis to the
simulations, we extracted the stability spectrum of the surface
from the structure factor data of Fig. \ref{figstruct}. For
an exponentially growing Fourier mode, we have 
$S(t) \propto \exp(2\omega t)$, and the growth rate can be
obtained as $\omega=\ln(S(t_2)/S(t_1))/2(t_2-t_1)$. 
Fig. \ref{figcomp} shows that the agreement with the theoretical
prediction is excellent, except for very small wave numbers.
This discrepancy can be explained by the fact that for large
wavelengths, more layers have to be included, as the diffusion
field in the vapor will be appreciably modified up to a
distance from the interface which is comparable to the wavelength.
In conclusion, the maximum growth rate and the corresponding
wave vector can be accurately predicted by a calculation
involving only a small number of layers.

For the example of Fig. \ref{figdroplet}, we find a maximum
growth rate of $\omega_s = 4.47 \times 10^{-4} w_0$ at a wave number 
$k_s = 0.349/a$. This surface mode grows more than four times
faster than the fastest bulk mode, and at a wavelength about
three times larger than the wavelength of the typical bulk pattern.
We verified indeed in our simulations that the ``incubation time'',
i.~e. the time after which the growing perturbations become
visible in a given visualization, was about four times larger
in the bulk than at the surface.

\begin{table}
\caption{Characteristic quantities for bulk and surface
modes in two dimensions for different temperatures}
\label{table1}
\begin{tabular}{ccccc}
\hline\noalign{\smallskip}
$kT/\epsaa$ & 0.4 & 0.5 & 0.6 & 0.7 \\
\noalign{\smallskip}\hline\noalign{\smallskip}
$\pavap$,$\pbvap$ & 0.0141 & 0.0354 & 0.0724 & 0.1412 \\
$\pasol$,$\pbsol$ & 0.4859 & 0.4646 & 0.4276 & 0.3588 \\
$\omega_b/w_0\times 1000$ & 0.016 & 0.106 & 0.255 & 0.008 \\
$\omega_s/w_0\times 1000$ & 0.383 & 0.447 & 0.342 & 0.003 \\
$\omega_s/\omega_b$ & 23.9 & 4.21 & 1.34 & 0.36 \\
$ak_b$ & 1.147 & 1.003 & 0.793 & 0.222 \\
$ak_s$ & 0.312 & 0.349 & 0.421 & 0.132 \\
$k_b/k_s$ & 3.68 & 2.87 & 1.88 & 1.68 \\
\noalign{\smallskip}\hline
\end{tabular}
\end{table}

\begin{table}
\caption{Same as table \ref{table1}, but in three dimensions.}
\label{table2}
\begin{tabular}{ccccc}
\hline\noalign{\smallskip}
$kT/\epsaa$ & 0.6 & 0.75 & 0.9 & 1.05 \\
\noalign{\smallskip}\hline\noalign{\smallskip}
$\pavap$,$\pbvap$ & 0.0141 & 0.0354 & 0.0724 & 0.1412 \\
$\pasol$,$\pbsol$ & 0.4859 & 0.4646 & 0.4276 & 0.3588 \\
$\omega_b/w_0\times 1000$ & 0.024 & 0.159 & 0.383 & 0.012 \\
$\omega_s/w_0\times 1000$ & 0.976 & 0.773 & 0.528 & 0.007 \\
$\omega_s/\omega_b$ & 40.7 & 4.86 & 1.37 & 0.61 \\
$ak_b$ & 1.453 & 1.259 & 0.986 & 0.272 \\
$ak_s$ & 0.414 & 0.410 & 0.505 & 0.204 \\
$k_b/k_s$ & 3.51 & 3.07 & 1.92 & 1.33 \\
\noalign{\smallskip}\hline
\end{tabular}
\end{table}
Having verified that our method gives accurate results, we can
apply it to study the behavior of bulk and surface modes for
different temperatures. The results are given in Table \ref{table1}.
All calculation were performed with $N=16$, except for the
highest temperature where $N=32$ was used. The bulk growth rate
has a maximum around $kT/\epsaa=0.6$ and decreases 
as the temperature is lowered, because the activated dynamics
lead to small values of the mobility for low temperatures.
On the other hand, for the highest temperature the system
is near a spinodal surface, and the driving force for
phase separation is small, leading also to a lower growth rate.
The surface growth rate follows the same trends.
At the highest temperature
considered, $\omega_s$ is lower than $\omega_b$, which
means that a proper surface mode does not exist any more.
The ratio $\omega_s/\omega_b$ increases monotonically
as the temperature is lowered. It should be noted
that the mean-field approximation becomes increasingly inaccurate for
low temperatures. Also, the interfaces become very sharp,
which leads to strong lattice effects.
 
It is also straightforward to repeat these calculations for
three dimensions. In fact, the phase diagram stays the same
if we scale the temperature by a factor of $1.5$.
On the other hand, the interface shapes and the growth rates
are different, because the balance between the discrete
Laplacians and the local terms is altered in Eqs. (\ref{chempots})
for the chemical potentials. In the stability calculations,
the magnitude of the stability matrix $\bf S$ is modified
whereas the terms arising from the Laplacians stay the same.
The final results for the growth rates and wave vectors, shown in
Table \ref{table2}, are similar to the two-dimensional case.

\subsection{Propagation of the surface patterns}
\nobreak
\noindent
The surface mode propagates
into the interior of the sample, enforcing its characteristic
wavelength in the direction parallel to the surface. This
behavior can be understood by considering the solution of the
eigenvalue equation away from the interface. Even if the
surface modes are localized at the vapor-mixture interface,
they are eigenmodes of the whole system and grow everywhere
with the same growth rate $\omega_s$. But away from the interface,
the initial state becomes homogeneous, and we can use
Eq. (\ref{omksym}), derived for the bulk, if we allow for
a decay of the amplitude in the direction normal to the
interface by introducing a complex wave number $k_x$. As the
growth rate and the $y$-component of the wave vector are
already known from the analysis of the surface instability,
$k_x$ is the only remaining unknown.
The relevant dispersion relation is $\omega_-(\vec k)$,
and we obtain the equation
\begeq
\Ak^2 \left(\epsaa-\epsab\right) - \Ak\left(S_{AA}-S_{AB}\right)
   + {\omega_s\over \bar M} = 0.
\endeq
This equation can either be solved numerically using the exact
expression for $\Ak$ or analytically with the approximation
$\Ak \approx -k_x^2-k_y^2$. The latter method leads to a
biquadratic equation for $k_x$. In both cases, there are four
solutions of the form
\begeq
k_x = \pm k' \pm i k''
\endeq
with $k'$ and $k''$ real and positive. The two solutions with
negative imaginary part diverge in the bulk and have to be 
discarded. The other two solutions lead to modes proportional
to $\exp(\pm ik'x + \omega_s t - k'' x)$, that is modes with an
oscillatory concentration profile and an envelope which decays
exponentially with the distance from the surface, but grows
exponentially in time. One can define a propagation velocity
of the decomposition front by the phase velocity of the envelope,
\begeq
v = {\omega_s\over k''}.
\endeq
We compared the values for the wavelength normal to the surface,
$\lambda_x = 2\pi/k'$, and the propagation velocity obtained by
this method to our simulations and found good agreement
\cite{Plapp97c}.

The competition between surface and bulk modes leads to the
interesting consequence that the thickness of the layer near
the surface where the surface-directed patterns prevail
depends on the strength of the initial fluctuations. This can
be seen as follows. For an exponentially growing mode with
initial amplitude $\delta_0$, the
time to reach a threshold amplitude $\delta_{nl}$ where
nonlinear couplings between different modes become important is
\begeq
\tau \sim {1\over \omega} \ln {\delta_{nl}\over \delta_0}.
\endeq
Now, a surface mode ``propagates'' into the sample approximately
between the time $\tau_s$ when it is well developed at the
surface, and the time $\tau_b$ when the bulk modes reach the
nonlinear stage. The distance $d$ the front propagates is
therefore given by
\begeq
d\sim v(\tau_b-\tau_s) = {1\over k''}
       \left({\omega_s\over\omega_b}-1\right)
       \ln{\delta_{nl}\over \delta_0}.
\label{thickscale}
\endeq
The thickness of the surface structures depends logarithmically
on the strength of the initial noise. This reasoning applies
both to the symmetrical checkerboard structures and the 
surface-directed spinodal waves and was confirmed by our 
simulations.

\section{Conclusions}
\nobreak
\noindent
We have developed mean field kinetic equations to describe the
dynamics of a lattice gas model of a binary alloy with vacancies
(ABv model) in which diffusion takes place by the vacancy
mechanism only. Despite the simplicity of the
model, we observe a rich variety of phase separation patterns
at free surfaces between an unstable mixture and a stable vapor.
The most spectacular effect is a fast surface mode. It creates
ordered patterns at the surface with a length scale which is
clearly distinct from the characteristic scale of the bulk
patterns. This mode appears in a small range of parameters
around a symmetric point where the mixture-vapor interface
is ``neutral'', that is none of the components of the mixture
segregates to the surface. On the other hand, if such segregation
occurs and is rapid enough, the surface mode is suppressed, and
instead surface-directed spinodal waves are observed which create
a striped pattern along the surface. Both patterns, once they
have formed at the surface, propagate into the sample over a
distance which is related to the difference in surface and bulk
growth rates and the strength of the initial fluctuations.

Our approach starts from a minimal model with very few assumptions.
Therefore, it cannot be used to model a particular experimental
situation. But it allows to identify some basic ingredients 
necessary for the formation of such surface structures in
a fairly well-defined setting, and we can draw some
conclusions which should be generally valid.

The existence of the fast surface mode is related to
the fact that in our model the surface atoms are more mobile
than in the bulk, an assumption which seems reasonable for
many interfaces between a dense and a dilute phase. The
characteristic growth rates of the surface modes can be
calculated by a linear stability analysis starting from the
initial interface profile. We have shown that accurate results
can be obtained by solving the resulting eigenvalue problem
in a small domain around the interface. The characteristics
of the propagation of the decomposition front can be obtained
by connecting the solution of the surface problem to the bulk
solution. This method could be applied as well to continuum
models of Ginzburg-Landau type by using an appropriate
discretization.

Which type of surface structure occurs is related to the
time scales for segregation and surface phase separation.
If the segregation is slow, surface spinodal decomposition
is dominant, otherwise spinodal waves are observed. As in
a realistic system, the interactions between different
species are always different, spinodal waves are the more
generic pattern. The surface mode should show up, however,
in a small range of initial compositions if the interactions
are not too different. More work is needed to clarify this
point.

A very interesting point is that the thickness of the layer 
of surface patterns depends on the initial fluctuation strength.
This is not the case in bulk spinodal decomposition,
where a rescaling of the initial fluctuations amounts simply to
a shift in time. 
The arguments leading to Eq. (\ref{thickscale}) are fairly
general and should apply to all systems where a fast surface
mode is in competition with a slow bulk mode.
In rapid quench experiments, the initial fluctuation
spectrum is mainly determined by the temperature {\em before}
the quench. Therefore, the thickness of the surface layer
may depend on the initial temperature.
In view of the logarithmic dependence of $d$ on the
noise strength, this effect might be difficult to observe;
however, if the initial state of the system is close to a
critical point, the fluctuation amplitude is a rapidly 
varying function of temperature.

Surface effects in spinodal decomposition have been studied
recently in polymers 
\cite{Jones91,Wiltzius91,Cumming92,Bruder92,Shi93,Harrison95}.
Evidently, our equations cannot be applied to polymers. But
from our results it is simple to construct a Ginzburg-Landau 
theory in which surface modes occur: it is sufficient to
introduce a mobility function which explicitly depends on the
density and which enhances diffusion on the surface.

It would be interesting to compare our findings to Monte Carlo 
simulations which naturally contain the fluctuations. It has
been shown, however, that short-range lattice gas models
do not exhibit linearly superposed, exponentially
growing modes in the early stages
of the phase separation. This can be attributed to the fact
that the initial noise is too strong for a linearization of
the equations of motion to be valid \cite{Binder84}. On the
other hand, for models with longer range interactions the
Cahn-Hilliard behavior is restored \cite{Laradji90}, and we
would expect surface modes in such models even with stochastic
dynamics.

In summary, our approach allows to explore the rich dynamics
of the ABv model and to relate the structures that form spontaneously
during phase separation to the parameters of the microscopic
model. Here, we have only explored a small part of the
possible behaviors in this model, because we have limited ourselves
to attractive interactions. There are other interesting questions
which could be addressed in the framework of this model and
using mean field kinetic equations, for example the influence
of the vacancy distribution on the coarsening behavior, or the
interplay between phase separation, short range ordering and 
vacancy distribution.

\acknowledgments

We have benefited from valuable discussions with W.~Dieterich, 
H.-P.~Fischer, and P.~Maass. We would like to thank Jean-Fran{\c c}ois
Colonna for his help with the generation of the 3D pictures.
One of us (M.P.) was supported by a grant from the Minist{\`e}re
de l'Enseignement Sup{\'e}rieur et de la Recherche (MESR).
Laboratoire de Physique de la Mati{\`e}re Condens{\'e}e is
Unit{\'e} de Recherche Associ{\'e}e (URA) 1254 to CNRS.


\appendix

\section*{Phase diagram}
\nobreak
\noindent
Let us first show that the Hamiltonian of the ABv-model
can be derived from a general ternary Hamiltonian. We assume
that the sites of the lattice can now be occupied by three
different sorts of atoms, $A$, $B$, or $C$. In terms of the
occupation numbers $n_i^A$, $n_i^B$, and $n_i^C$, the energy
of a configuration is given by
\begeq
H = - \sum_{\alpha, \beta} \sum_{\ij} 
  \epsilon'_{\alpha\beta} n_i^\alpha n_j^\beta
  - \sum_\alpha \sum_{i=1}^N {\mu'}_0^\alpha n_i^\alpha.
\endeq
The interaction energies in the ternary system are primed to
distinguish them from the $\epsilon$'s in Eq. (\ref{hamilabv}).
Here, the ${\mu'}_0^\alpha$ are external chemical potentials,
equivalent in the spin language to ``generalized magnetic fields''
acting only on one possible spin state. Using the constraint
of single occupancy, $n_i^A+n_i^B+n_i^C=1\,\forall i$, we can
eliminate the occupation numbers of one species, say $C$, and
we obtain
\begeq
H = -\sum_{\ij} \left[\epsaa n_i^A n_j^A 
   +\epsbb n_i^B n_j^B
   +\epsab(n_i^A n_j^B + n_i^B n_j^A)\right]
    - \sum_{i=1}^N\left(\mu^A_0 n_i^A + \mu^B_0 n_i^B\right),
\label{equiveq}
\endeq
where the effective interaction energies and chemical potentials
appearing in this last equation are
\begin{eqnarray}
\epsaa & = & \epsaa' + \epsilon_{CC}' - 2\epsilon_{AC}', \\
\epsbb & = & \epsbb' + \epsilon_{CC}' - 2\epsilon_{BC}', \\
\epsab & = & \epsab' + \epsilon_{CC}' - \epsilon_{AC}' - \epsilon_{BC}', \\
\mu^A_0 & = & {\mu'}^A_0 - {\mu'}^C_0 - z\epsilon_{CC}' + 
     z\epsilon_{AC}'\quad{\rm and} \\
\mu^B_0 & = & {\mu'}^B_0 - {\mu'}^C_0 - z\epsilon_{CC}' + z\epsilon_{BC}'.
\end{eqnarray}
When the total number of particles of
each species is conserved, the chemical potential terms
in Eq. (\ref{equiveq}) are constants. Then, we are back
to the Hamiltonian Eq. (\ref{hamilabv}).

\begin{figure}
\centerline{
 \psfig{file=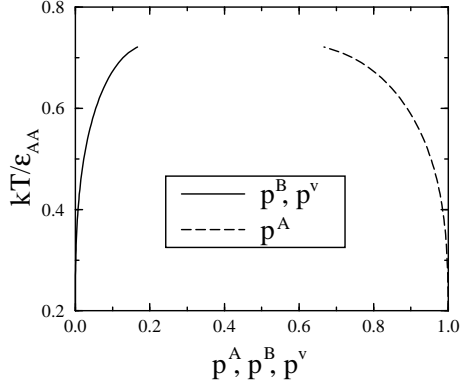,width=0.4\textwidth}}
\caption{Phase diagram of the completely symmetric ABv-model.
We show the equilibrium concentrations in the $\alpha$ (A-rich)
phase for three-phase coexistence. The concentrations of the
other equilibrium phases can be obtained by symmetry.
The representative points of the three phases in equilibrium
give the central triangle in the diagram in Fig. \ref{figphasediag}.}
\label{figconc}
\end{figure}
Let us now consider the three state Potts model, with ternary
interaction energies $\epsaa'=\epsbb'=\epsilon_{CC}'=1$ and
$\epsilon_{\alpha\beta}'=0$ for $\alpha\neq\beta$. This 
gives the effective interactions $\epsaa=\epsbb=2,\epsab=1$;
by changing the temperature scale, we obtain the values
$\epsaa=\epsbb=1,\epsab=0.5$ used in the present paper.
Furthermore, from the symmetry of the Potts model it is obvious
that three phase coexistence is possible below the critical
temperature at zero magnetic fields. This is equivalent, in the
ABv model, to
\begeq
\mu^A = -z\epsaa p^A -z\epsab p^B + kT{p^A\over 1-p^A-p^B} = -z\epsaa/2
\endeq
and an equivalent equation for $\mu^B$. These equations have to be
solved numerically to obtain the equilibrium concentrations. The
result for the A-rich phase is shown in Fig. \ref{figconc}. The 
coexistence of A-rich, B-rich, and ``vapor''-phases terminates at a 
quadruple point: a fourth minimum in the free energy surface, located at the
symmetric point $p^A=p^B=p^v=1/3$, develops, and at $kT=1/(2\ln 2)\approx0.721$
the four minima are exactly at the same level. Above this temperature,
there is a narrow temperature range above which the four minima persist,
but now the fourth minimum is lowest, and we have three different
regions of three-phase coexistence (not shown in the figure). Above
$kT=0.75$, only the ``symmetric'' minimum remains.



\begin{thebibliography}{99}
\bibitem{Note} Note that in Ref.~\cite{Plapp97c}, we have used
  no-flux boundary conditions at the limits of the solution
  region, and obtained quite precise values already with four
  layers. This method being more cumbersome to implement, we
  did not carry out a more systematic study.
\end{thebibliography}
\end{document}